\documentclass[a4paper,11pt]{article}
\usepackage[utf8]{inputenc}
\usepackage[T1]{fontenc}
\usepackage[UKenglish]{isodate}

\usepackage{mathrsfs}
\usepackage{bm}
\usepackage{bbm}
\usepackage{palatino}

\usepackage[margin=2cm]{geometry}
\usepackage{mathtools}
\usepackage[usenames,dvipsnames]{xcolor}
\usepackage[pdftex]{graphicx}
\usepackage{amsmath,amssymb,amsfonts}

\usepackage{amsthm}
\usepackage[textsize=small,textwidth=1.3in]{todonotes}
\usepackage[english]{babel}

\usepackage[colorlinks=true,urlcolor=Mahogany,linkcolor=Mahogany,citecolor=Mahogany,plainpages=false,pdfpagelabels]{hyperref}

\usepackage{authblk}

\usepackage{multirow}

\usepackage{tikz}
\usepackage{verbatim}
\usetikzlibrary{shapes.geometric,plotmarks,backgrounds,fit,calc,circuits.ee.IEC}
\usetikzlibrary{decorations.pathreplacing}
\usepackage[backend=bibtex,maxcitenames=4,maxalphanames=100,maxbibnames=100,isbn=false]{biblatex}
\tikzset{phase/.style = {draw,fill,shape=circle,minimum size=5pt,inner sep=0pt},crossx/.style={path picture={ 
\draw[thick,black,inner sep=0pt]
(path picture bounding box.south east) -- (path picture bounding box.north west) (path picture bounding box.south west) -- (path picture bounding box.north east);
}}, cross/.style={path picture={ 
\draw[thick,black](path picture bounding box.north) -- (path picture bounding box.south) (path picture bounding box.west) -- (path picture bounding box.east);
}}, not/.style={draw,circle,cross,minimum width=0.3 cm}}
\usepackage{xcolor}

\usepackage{color, colortbl}
\usepackage{multirow}
\usepackage{easybmat}
\usepackage{physics}
\usepackage{epstopdf}

\usepackage{cleveref} 

\usepackage{lscape} 

\usepackage{biblatex}
\usepackage{caption} 
\usepackage{subcaption}

\usepackage{tabularx} 

\usepackage{widetext}

\newtheorem{theorem}{Theorem}
\newtheorem{lemma}[theorem]{Lemma}

\newtheorem{proposition}[theorem]{Proposition}
\newtheorem{definition}[theorem]{Definition}

\newtheorem{example}[theorem]{Example}

\providecommand{\keywords}[1]{\textbf{\textit{Keywords:}} #1}

\long\def\symbolfootnote[#1]#2{\begingroup\def\thefootnote{\fnsymbol{footnote}}\footnote[#1]{#2}\endgroup}

\newcounter{protocol}

\begin{document}

\title{\vspace{-1.3cm}Stabilizer codes for Open Quantum Systems}
\author[1,2]{Francisco Revson F. Pereira\footnote{revson.ee@gmail.com; current affiliation: IQM,
Nymphenburgerstr. 86, 80636 Munich, Germany.}}
\author[1,2]{Stefano Mancini}
\author[3]{Giuliano G. La Guardia}
\affil[1]{School of Science and Technology, University of Camerino, I-62032 Camerino, Italy}
\affil[2]{INFN, Sezione di Perugia, I-06123 Perugia, Italy}
\affil[3]{Department of Mathematics and Statistics, State University of Ponta Grossa, 84030-900 Ponta
Grossa, PR, Brazil\vspace{-1cm}}

\date{}

\maketitle              

\begin{abstract}
The Lindblad master equation describes the evolution of a large variety of open quantum systems.
An important property of some open quantum systems is the existence of decoherence-free subspaces. 
A quantum state from a decoherence-free subspace will evolve unitarily. However, there is no procedural and 
optimal method for constructing a decoherence-free subspace.
In this paper, we develop tools for constructing decoherence-free stabilizer codes for open quantum systems governed by 
Lindblad master equation. This is done by pursuing an extension of the stabilizer formalism beyond the
celebrated group structure of Pauli error operators. We then show how to utilize decoherence-free stabilizer codes
in quantum metrology in order to attain the Heisenberg limit scaling with low computational complexity.

\newline
\end{abstract}


\section{Introduction}
\label{sec:Introduction}
\noindent

The second quantum revolution emerges from the possibility of designing and controlling quantum systems. 
The complexity of controlling quantum systems can be reduced by decreasing the noise due to system-environment 
interaction. 
This can be achieved by resorting to quantum error correcting codes. Among them are the stabilizer codes~\cite{GottesmanThesis}.
These codes were proposed by Gottesman~\cite{GottesmanThesis}. Several works have extended the original construction method in order
to incorporate Hilbert spaces and quantum systems with different
structures~\cite{Calderbank1998,Gottesman2001,Ollivier2003,Brun2006,Ketkar2006,Cafaro2010,Cafaro2012,Galindo2019,Noh2020}.
As an important result, it was shown that stabilizer codes exist if and only if there exist classical block codes obeying an orthogonality rule;
e.g., Euclidean and Hermitian self-orthogonality~\cite{Ketkar2006}. Such a duality between stabilizer codes and classical block codes has stimulated several
works~\cite{Ketkar2006,Cafaro2010,Cafaro2012,Guardia2017,Galindo2019,Pereira2019,Pereira2022,Noh2020,Pereira2021}.
Nevertheless, there is still room for novelties. In particular, we can find at least two relevant points not envisioned in previous works.

Firstly, consider the error set and the algebraic structure associated with it. One will see that the errors obey a group structure with the composition of
operators as the group operation. However, in several practical cases, one has an additional operation in play, which is the sum of operators. In these 
applications, a ring or vector space structure is needed. This is the framework we are going to consider in this paper.
As a consequence of this new and intricate formulation, the standard dual structure of stabilizer codes will not be block codes in the general case.
We need to abandon the idea of classical codes as vector spaces and work with additive groups.
For these additive groups, we introduce a new sum operation that corresponds to the sum and composition of operators.
A homomorphism is given between operators and additive codes.

Secondly, stabilizer codes are often designed for a specific quantum channel, or anyway their performance vary from channel
to channel~\cite{Cafaro2011}.
Having a dynamical evolution means to deal with time varying Kraus operators, or equivalently with time varying quantum channels.
Hence, in such a case it might be not satisfactory to resort to the standard stabilizer codes construction.
In this paper, we consider an open quantum system described by a Lindblad master equation.
This class of equations is the most general form for the generator of a quantum dynamical semigroup.
We construct stabilizer codes able to eliminate the dissipator part of the Lindblad master equation, thus turning
the evolution into unitary. As we are going to show, this is possible since the stabilizer code corresponds to a decoherence-free subspace.
A state from a decoherence-free subspace will evolve unitarily; i.e., the dissipator part of
the Lindblad master equation will not contribute to the evolution of the state \cite{Karasik2008}.
Even though the stabilizer code constructed is a subspace of the corresponding decoherence-free subspace,
an important advancement is made here. Using the stabilizer code construction we can derive a procedural and optimal method, in terms of
computational complexity, for constructing the decoherence-free subspace that corresponds to the stabilizer code.

Decoherence-free subspace can be regarded as a particular case of subsystem codes~\cite{Bacon2006}. Suppose we decompose the Hilbert space as
$\mathcal{H} = (\mathcal{C}\otimes\mathcal{D})\oplus\mathcal{B}$, where the subsystem code belongs to $\mathcal{C}$, $\mathcal{D}$ is a second subsystem, and
we are partitioning $\mathcal{H}$ into two subspaces, $\mathcal{B}$ and $\mathcal{C}\otimes\mathcal{D}$. The concept of a noiseless subsystem is that symmetries
on the system-environment evolution factor the interaction Hamiltonian with respect to some subsystem structure on the
Hilbert space $\mathcal{H}$~\cite{Bacon2006}. One can describe decoherence-free subspaces as a subsystem code where there is no
subsystem $\mathcal{D}$ and the code relies on symmetries that factor the interaction Hamiltonian with respect to system and environment.

Quantum metrology aims at using quantum systems in order to improve the estimation of parameters.
It is possible to show that the reduction one can obtain in terms of the number of probes a quantum system can achieve is 
quadratically faster than the best classical strategy~\cite{Giovannetti2006}. This is characterized by the Heisenberg limit (HL). 
In this paper, as an application of the developed theory, we present a method to achieve the HL using decoherence-free stabilizer codes.
It differs quite significantly from the existing methods in the literature. In previous works, under the hypothesis that the Hamiltonian is
not in Lindblad span (HNLS),
joint encoding and decoding schemes are presented in order to obtain the Heisenberg limit
\cite{DemkowiczDobrzanski2017,Zhou2018,Layden2019,Gorecki2020,DemkowiczDobrzanski2020}. 
In our method, using decoherence-free subspaces for quantum metrology, the dissipator part of the open quantum system does not contribute to the evolution.
Hence, under certain hypothesis, we demonstrate that it is possible to achieve the HL using decoherence-free stabilizer codes.
Furthermore, since the system evolves unitarily without any interaction between
the system and environment, there is no need for a decoding process. Therefore, we are free of any further errors that could arrive in the decoding process. 
It is also clear that the computational complexity is reduced when compared with other schemes.

This paper is organized as follows. In Section~\ref{sec:Preliminaries} we present the preliminary concepts used to
elaborate the results in this paper. The mathematical tools needed to construct the stabilizer codes
for noise operators having a tensor product description and a more general description
are shown in Sections~\ref{sec:Particular_main}~and~\ref{sec:Main}, respectively. A connection between stabilizer codes and decoherence-free subspaces is made.
We also illustrate some of the new structures by presenting several examples.
Next, Section~\ref{sec:channelEstimation} demonstrates the applicability of the stabilizer codes in the area of
quantum metrology. A condition for probing a quantum system using stabilizer codes in order to obtain the 
Heisenberg limit scaling is stated and analyzed.
The final remarks and future perspectives are given in Section~\ref{sec:Conclusion}.


\section{Preliminaries}
\label{sec:Preliminaries}
In this section, we review some formulations needed to understand the main results of the paper.
Firstly, we review the evolution of open quantum systems and describe the idea of
decoherence-free subspace in Lindblad master equations. Secondly,
the stabilizer formalism is presented. We focus on the main aspect considered in this paper,
which is the algebraic structure obeyed by the errors. The first result of the paper is also given,
connecting decoherence-free subspaces and stabilizer codes.

\subsection{Open Quantum Systems and Decoherence-free Subspace}
\label{sec:cyclicCodes}

In this paper we deal with open quantum systems evolving by means of a Markovian master equation. Suppose
$q$ is a prime power and let $S$ be a system  modeled by a $q$-dimensional Hilbert
space $\mathcal{H}_S$ (we also use the notation $\mathcal{H}_q$ when it is clear which system is considered) interacts with a reservoir system $R$ associated with a Hilbert space $\mathcal{H}_R$.
Then the full Hamiltonian can be decomposed as
\begin{equation}
H = H_S\otimes\mathbb{I} + \mathbb{I}\otimes H_R + H_{\text{int}},
\end{equation}with $H_S$, $H_R$, and $H_{\text{int}}$ the system, reservoir, and interaction Hamiltonians, respectively,
and $\mathbb{I}$ is the identity operator. Instead of analyzing the full system-reservoir evolution, we are only interested
in the evolution of a reduced system density operator $\rho$. This is obtained by tracing over the reservoir in the
full density operator. As a result, the dynamics of the reduced density operator $\rho$ is given by~\cite{HeinzPeterBreuer2007}
\begin{equation}
\frac{\partial\rho}{\partial t} = -i[H_S,\rho] + L_D(\rho),
\label{Eq:masterEquation}
\end{equation}where
\begin{equation}
L_D(\rho) = \frac{1}{2}\sum_{l=1}^M \lambda_l ([J_l, \rho J_l^\dagger] + [J_l\rho,J_l^\dagger])
\end{equation}is the decoherence evolution originated from the system-reservoir coupling,
with $M\leq N^2 - 1$ where $N$ is the number of qudit systems, $q = d^N$, and $\{J_l\}_{l=1}^M$
are the Lindblad operators. We call this part of the evolution throughout the paper as the dissipator part.

\begin{definition}\cite{Karasik2008}
      Let the time evolution of an open quantum system with Hilbert space $\mathcal{H}_S$ be described by means of a Markovian
      master equation and $D(\mathcal{H})$ be the set of density matrices of a given Hilbert space $\mathcal{H}$. Then a decoherence-free subspace (DFS) $\mathcal{H}_\text{DFS}$ of $\mathcal{H}_S$ is such that all pure
      states $\rho(t)\in D(\mathcal{H}_\text{DFS})$ satisfies
      \begin{equation}
            \frac{d \text{Tr}\{\rho^2(t)\}}{dt} = 0, \forall t\geq 0,\qquad\text{ with }\qquad\text{Tr}\{\rho^2(0)\}=1.
      \end{equation}On the other hand, a subspace $\mathcal{H}_\text{sDFS}$ is called \emph{strong} decoherence-free subspace (sDFS) if for all
      pure $\rho(t)\in D(\mathcal{H}_\text{sDFS})$ one has
      \begin{equation}
            L_D(\rho(t)) = 0, \qquad \text{ and } \qquad \rho^2(t) = \rho(t), \forall t.
      \end{equation}
      \label{Def:DFSsDFS}
\end{definition}

From Definition~\ref{Def:DFSsDFS}, it is clear that strong DFS is a sufficient but not necessary
condition to have a DFS.


\begin{proposition}\cite[Theorem 4, Proposition 5, Theorem 6]{Karasik2008}
      Let the time evolution be given by the Markovian open system dynamics shown in Eq.~(\ref{Eq:masterEquation}).
      Assume $\Gamma = \sum_{l=1}^M \lambda_l J_l^\dagger J_l$.
      On the one hand, the space $\mathcal{P} = \text{span}\{\ket{\psi_i}\}_{i=1,\ldots,K}$ is a DFS for all time $t$
      if and only if $J_l\ket{\psi_k} = c_l\ket{\psi_k}$, for all $l=1,\ldots,M$ and $k = 1,\ldots, K$, and
      the commutator $[H_{ev},J_l]$ has eigenvalues equal to zero for all $\ket{\psi_k}\in\mathcal{P}$, and $l=1,\ldots,M$. Here
      \begin{equation}
            H_{ev} = H_S + \frac{i}{2}\sum_{l = 1}^M \lambda_l(c_l^* J_l - c_l J_l^\dagger).
      \end{equation}
      On the other hand, $\mathcal{P}$ is sDFS if the
      commutators $[H_S,J_l]$, for $l=1,...,M$, and $[H_S,\Gamma]$ have
      eigenvalues equal to zero for all $\ket{\psi_k}\in\mathcal{P}$. Furthermore, $J_l\ket{\psi_k} = c_l\ket{\psi_k}$,
      and $\Gamma\ket{\psi_k} = g\ket{\psi_k}$, for all $l=1,\ldots,M$ and $k = 1,\ldots, K$, where
      $g = \sum_{l=1}^M\lambda_l|c_l|^2$.
      \label{Proposition:DFS}
\end{proposition}

We are going to show below a case study in which the operators in the dissipator part of the evolution cannot
be described
in a tensor product form of Pauli operators. Later, we will show that a stabilizer code can be derived from the (strong) DFS,
but one cannot rely on the standard stabilizer formalism, which will be explained in the next subsection, for such codes.

\begin{example}
      Consider an $N$-qubit quantum system with the dynamics described by the master equation
      \begin{equation}
            \frac{\partial\rho}{\partial t} = -i[H_S,\rho] + \frac{\gamma}{2}(2J\rho J^\dagger - J^\dagger J\rho - \rho J^\dagger J),
            \label{Ex7}
      \end{equation}with
      \begin{equation}
            J = \sum_{j=1}^N J_j \qquad\text{ and }\qquad J_j = \frac{s+c}{2}(\sigma_{xj} + i\sigma_{yj} + \sigma_{zj}),
      \end{equation}where $\sigma_{xj}, \sigma_{yj}, \sigma_{zj}$ are the Pauli operators on the $j$-th qubit,
      $s = \sinh(r)$, $c=\cosh(r)$, and $r$ is the (real) squeezing parameter
      derived from the assumption that the reservoir is given by a squeezed vacuum state.
      For $r\neq 0$, the eigenvectors of $J$ are all possible tensor products of $\ket{\psi_{+}}$ and $\ket{\psi_{-}}$, where
      $\ket{\psi_{+}} = \ket{0}$ and $\ket{\psi_{-}} = \frac{\ket{0} - \ket{1}}{\sqrt{2}}$. For each $J_j$, the eigenvalue of $\ket{\psi_{\pm}}$ is equal to
      $\pm(\frac{s+c}{2})$. Thus, if the number of $\ket{\psi_{+}}$ in the eigenvector of $J$ is $n_{+}$, then the eigenvalue of such
      eigenvector is $(n_{+} - n_{-})\frac{s+c}{2}$, where $n_{-} = N - n_{+}$.
      Notice that it is not possible to describe $J$ as only tensor products of Pauli operators. This forbid us to use standard constructions
      of stabilizer codes in the theory that follows. Therefore, before presenting our main results, we need to extend the stabilizer formulation
      to a suitable type of errors, in conjunction with the operations allowed between them.
      Now, to compute the expression for $H_S$ so that there exists a nonempty (strong) DFS of the
      eigenspace of $J$, we use the commutativity condition on $H_S$ given in Proposition~\ref{Proposition:DFS}. From it, one possible solution is to
      have the system Hamiltonian with the form
      \begin{equation}
            H_S = \sum_{j = 1}^N\frac{\gamma}{4}(n_+ - n_{-})(s+c)^{2}\sigma_{yj},
      \end{equation}where $n_+ \neq n_{-}$. Observe that the choice of $H_S$ implies that $H_{ev} = 0$ and, therefore, the space spanned by
            $\ket{\psi_{+}}$ and $\ket{\psi_{-}}$ is sDFS.
      \label{Example1}
\end{example}

\subsection{Stabilizer Codes}
\label{sec:stabilizerCodes}

The theory of stabilizer codes, introduced by Gottesman, has been long studied, analyzed, and
extended~\cite{Gaitan2008,Mancini2020,LidarBrunBook2014}. However, the physical environment over
which the stabilizer code will be used is commonly ignored, relying on the error-correction capability only
over the minimal distance of the code. Before presenting the approach taken in this paper, we recall the stabilizer formalism. Actually, we refer to the stabilizer formalism introduced by Gottesman, the connection between
the stabilizer group and associated additive code by standard stabilizer formalism.

A stabilizer code $\mathcal{Q}$ is a subspace of a $N$-qubit system described by $\mathbb{C}^{2^N}$ stabilized by the elements
of an abelian subgroup $S$ of the error group $G_N$ over $N$ qubits.
We can mathematically describe a stabilizer code as
\begin{equation}
\mathcal{Q} = \bigcap_{E\in S}\{\ket{\psi}\in\mathcal{H}_q\colon E\ket{\psi} = \ket{\psi}\}.
\end{equation}For the characterization of correctable and uncorrectable errors, we need to introduce the concept
of centralizer of a subgroup of $G_N$ and center of $G_N$. The subgroup $C_{G_N}(S)$ of $G_N$, given by
\begin{equation}
C_{G_N}(S) = \{E\in G_N\colon E F = F E \text{ for all }F\in S\},
\end{equation}is called the centralizer of $S$ in $G_N$. Observe that the commutativity property of $S$ implies
$S\leq C_{G_N}(S)$. The center of $G_N$, denoted by $Z(G_N)$, is the subgroup $Z(G_N) = C_{G_N}(G_N)$.
The group $SZ(G_N)$ is given by the elements $Ez$, where $E\in S$ and $z\in Z(G_N)$.
The following lemma characterizes the relation among
correctable errors, stabilizer, centralizer, and the center of a
quantum code.

\begin{lemma}\cite[Lemma  11]{Ketkar2006}
Let $S\leq G_N$ be the stabilizer group of a stabilizer code $\mathcal{Q}$ of dimension greater than one. An error
$E\in G_N$ is detectable by the stabilizer code $\mathcal{Q}$ if and only if $E$ is an element of $SZ(G_N)$ or $E$ does not
belong to the centralizer $C_{G_N}(S)$.
\label{Lemma:Stabilizer}
\end{lemma}

An important achievement of standard stabilizer formalism is showing the
equivalence of stabilizer codes and additive classical codes. For a
$K$-dimensional quantum code with minimum distance $d$ and living in the
Hilbert space $\mathcal{H}_q^{\otimes N}$; i.e., with parameters $((N,K,d))_q$, there
exists an additive classical code $C$ (and its symplectic dual $C^{\perp_s}$) which is
equivalent to the stabilizer code. As shown in Proposition~\ref{Ketkar:Thm13},
the minimum distance of the stabilizer code is computed from the symplectic weight.
We define the symplectic weight of a vector $(\bm{a}|\bm{b})\in\mathbb{F}_q^{2N}$ as
$\text{swt}((\bm{a}|\bm{b})) = |\{k\colon (a_k,b_k)\neq (0,0)\}|.$
One of the main results of this paper is the extension the equivalence between classical codes
and quantum stabilizer codes to more general errors, in particular non-Pauli errors. Due to
the complex algebraic structure describing the errors, we show sufficient
conditions for a classical code to be equivalent to a stabilizer code.

\begin{proposition}\cite[Theorem  13]{Ketkar2006}
An $((N, K, d))_q$ stabilizer code exists if and only if there exists
an additive code $C \subseteq \mathbb{F}_q^{2N}$ of size $|C| = q^N/K$
such that $C \subseteq C^{\perp_s}$ and $\text{swt}(C^{\perp_s}\setminus C) = d$
if $K > 1$ (and $\text{swt}(C^{\perp_s}) = d$ if $K = 1$), where $\text{swt}$
denotes the symplectic weight, and the symplectic weight of a set is the
minimum symplectic weight of the elements in the set.
\label{Ketkar:Thm13}
\end{proposition}

Now, we are going to describe in detail the error basis and error vector space used through the paper.
A set $\mathcal{E}$ of operators on $\mathbb{C}^2$ is denoted a
\emph{nice error basis} if it attains three conditions: (a) it contains the identity operator, (b) it is closed
under the composition of operators, (c) the trace $\text{Tr}\{A^\dagger B\} = 0$ for distinct elements $A,B\in\mathcal{E}$.
In this paper, we consider the error basis
\begin{equation}
      \mathcal{E} = \{\mathbb{I}, \sigma_x, \sigma_y, \sigma_z\},
\label{Eq:12}
\end{equation}where $\mathbb{I}$ is the identity operator and $\sigma_i$, for $i=1,2,3$,
are the Pauli matrices. The inner product of two distinct elements $A,B$ in $\mathcal{E}$ is given by
\begin{equation}
      \langle A,B\rangle = \text{Tr}\{A^\dagger B\}.
\end{equation}Clearly, $\mathcal{E}$ is a nice error basis. Additionally, we have that if $\mathcal{E}_1$ and
$\mathcal{E}_2$ are nice error bases, then $\mathcal{E}^2 = \{\sf{E_1}\otimes \sf{E_2}\colon E_1\in\mathcal{E}_1,E_2\in\mathcal{E}_2\}$
is a nice error basis as well.

\begin{proposition}
      Let $A = a_0\mathbb{I} + a_1\sigma_x + a_2\sigma_y + a_3\sigma_z$ and
      $B = b_0\mathbb{I} + b_1\sigma_x + b_2\sigma_y + b_3\sigma_z$ be two elements generated by $\mathcal{E}$.
      Then
      \begin{equation}
      [A,B] = 2i\Big{(}(a_2 b_3 - b_2 a_3)\sigma_x + (a_3 b_1 - b_3 a_1)\sigma_y + (a_1 b_2 - b_1 a_2)\sigma_z\Big{)}.
      \end{equation}
      \label{Prop:CommutationRelation}
\end{proposition}

\begin{proof}
      It follows from the commutation relations of the Pauli operators 
      $[\sigma_i,\sigma_j] = 2i\epsilon_{ijk}\sigma_k$, for $i,j,k = 1, 2, 3$.
\end{proof}

\begin{definition}
Let $\mathcal{E}^N$ be the error basis constructed
as $N$-fold tensor product of the Pauli matrices shown in Eq.~(\ref{Eq:12}). The error group, denoted by $G_N$,
is the vector space over $\mathbb{C}$ consisting of elements in $\mathcal{E}^N$.
\label{Def:GN}
\end{definition}

For explicit examples of the error vector space $G_N$, see Section~\ref{sec:Main}.

In the definition of error set and operations allowed among its elements, there is
a significant turn concerning the one utilized in standard stabilizer formalism.
There, only the composition of operators is considered. Here, we also have the sum
of operators, thus obtaining a vector space structure for the error set.
Notice, however, that to define the stabilizer set in both approaches (the
standard one and the one used in this paper) only composition of operators
is necessary, besides the commutativity of its elements.

Suppose the evolution of a state $\rho(t)$ is given by a Markovian master equation
with dissipator part described by operators from the set $\mathcal{J'} = \{J_l\colon l=1,\dots, M\}$.
Add the operator $\Gamma = \sum_{l=1}^M \lambda_l J_l^\dagger J_l$ to the set $\mathcal{J'}$
so that $\mathcal{J} = \mathcal{J'}\cup \Gamma$.
Assume the existence of a DFS or sDFS satisfying the assumptions of Proposition~\ref{Proposition:DFS}.
Then we can construct the following stabilizer sets
\begin{equation}
      \mathcal{S}_\text{sDFS} := \langle S_1, \ldots, S_{M+1}\colon S_l = c_l^{-1} J_l, \text{ for }l=1,\ldots,M+1,\text{ where } J_l\in\mathcal{J}\rangle,
      \label{Eq:StabilizerGroup}
\end{equation}or
\begin{equation}
      \mathcal{S}_\text{DFS} := \langle S_1, \ldots, S_{M}\colon S_l = c_l^{-1} J_l, \text{ for }l=1,\ldots,M,\text{ where }J_l\in\mathcal{J'}\rangle.
      \label{Eq:StabilizerGroup2}
\end{equation}
Suppose $\mathcal{Q}$ is the joint eigenspace with eigenvalue $+1$ for every element in $\mathcal{S}_\text{sDFS}$; i.e.,
$\mathcal{S}_\text{sDFS}$ stabilizes $\mathcal{Q}$. If $[S_i,S_j] = 0$, for all $i,j=1,\ldots,M+1$, then
$\mathcal{S}_\text{sDFS}$ is an abelian group.
Furthermore, if the system Hamiltonian $H_S$ belongs to the centralizer $C_{G_N}(\mathcal{S}_\text{sDFS})$, then we can conclude from
Proposition~\ref{Proposition:DFS} that $\mathcal{Q}$ is sDFS.
Similar arguments can be used for DFS, where the stabilizer group is given by
$\mathcal{S}_\text{DFS}$ and the commutativity condition is imposed over $H_{ev}$.
A stabilizer code that is also (sDFS) DFS will be called a (strong) decoherence-free stabilizer code.

\begin{theorem}
      Let the time evolution of the quantum system be given by the Markovian master equation shown in
      Eq.~(\ref{Eq:masterEquation}).
      Suppose there exists a nontrivial maximal joint $+1$-eigenspace $\mathcal{Q}$ of the abelian group
      $\mathcal{S}_\text{sDFS}$ or $\mathcal{S}_\text{DFS}$ constructed in Eq.~(\ref{Eq:StabilizerGroup})
      and (\ref{Eq:StabilizerGroup2}), respectively.
      If
      \begin{enumerate}
            \item $H_S + \frac{i}{2}\sum_{l = 1}^M \lambda_l(c_l^* J_l - c_l J_l^\dagger)$ belongs to $C_{G_N}(\mathcal{S}_\text{DFS})$,
            then $\mathcal{Q}$ is a stabilizer code and a decoherence-free subspace;
            \item $H_S$ belongs to $C_{G_N}(\mathcal{S}_\text{sDFS})$ and $S_{M+1}\in\mathcal{S}_\text{sDFS}$ stabilizes $\mathcal{Q}$, then
            $\mathcal{Q}$ is a stabilizer code and a strong decoherence-free subspace.
      \end{enumerate}
      We call $\mathcal{Q}$ a (strong) decoherence-free stabilizer code.
      \label{Theorem:DFStoStabilizer}
\end{theorem}

\begin{proof}
      Consider Claim $1$. First of all, notice that the claim $\mathcal{Q}$ is a stabilizer
      code of $\mathcal{S}_\text{DFS}$ follows from the fact that $\mathcal{Q}$ is the nontrivial maximal
      $+1$-eigenspace of $\mathcal{S}_\text{DFS}$. Secondly, for any $\ket{\psi}\in\mathcal{Q}$
      and $S_l\in\mathcal{S}_\text{sDFS}$ we have
            \begin{equation}
                  J_l\ket{\psi} = c_l S_l\ket{\psi} = c_l\ket{\psi}.
                  \label{Eq:ProofThm7}
            \end{equation}Since $H_S + \frac{i}{2}\sum_{l = 1}^M \lambda_l(c_l^* J_l - c_l J_l^\dagger)$
      belongs to $C_{G_N}(\mathcal{S}_\text{DFS})$, then the commutator of
      $H_S + \frac{i}{2}\sum_{l = 1}^M \lambda_l(c_l^* J_l - c_l J_l^\dagger)$ with any element in
      $\mathcal{S}_\text{DFS}$ has eigenvalue equal to zero. Therefore, from Eq.~(\ref{Eq:ProofThm7}) and
      Proposition~\ref{Proposition:DFS}, we have that $\mathcal{Q}$ is also a decoherence-free subspace.

      Claim $2$ follows the same reasoning.
\end{proof}

A connection between stabilizer codes and decoherence-free subspace is given in Theorem~\ref{Theorem:DFStoStabilizer}.
Differently from previous works, such as Ref.~\cite{Lidar1999}, we give a direct algebraic relation
between the Lindblad operators, DFSs, and stabilizer codes. It is shown in Ref.~\cite{Lidar1999} that
DFSs are a specific class of quantum error correcting codes, but no constructive method to derive the stabilizer set
from the Lindblad operators is shown. Furthermore, as will be shown in the following sections, we extend the stabilizer
description to classical error-correcting codes defined over the complex number field.
More precisely, the standard theory of quantum error-correcting code contains quantum codes derived from classical codes,
i.e., codes defined over finite fields. In this new context, we consider classical codes defined
over $\mathbb{C}$, the complex field which has characteristic zero, and this fact modifies completely the techniques to
be applied in the constructions of our results.
To the best of the author's knowledge,
this is the first work presenting such a formulation.
In particular, there are DFS that have a stabilizer code as subspace. This inclusion may or may not be proper.
However, dealing with stabilizer codes can produce results that we could not obtain otherwise. In fact, one can find encoding
methods for stabilizer codes that are procedural and optimum algorithms for creating the corresponding code space.
Additionally, set membership can be optimally implemented by decoding methods.
Later in the paper we construct an algorithm for quantum metrology that uses set membership as one of the
important steps.
Therefore, dealing with decoherence-free stabilizer code instead of the whole decoherence-free
subspace is computationally relevant for several applications.

The connection between decoherence-free subspaces and stabilizer codes is expanded in the following two sections.
Firstly, errors with a particular structure are considered. The considered structure simplifies the stabilizer
formalism and the connection between stabilizers and classical codes.
Afterwards, the restriction is relaxed and generalized errors are considered.

\section{Decoherence-Free Stabilizer Codes for Tensor-Product Noise}
\label{sec:Particular_main}
We have seen necessary and sufficient conditions for a subspace
$\mathcal{P}\subseteq\mathcal{H}_S$ to be decoherence-free or strong decoherence-free in Proposition~\ref{Proposition:DFS}.
Additionally, we established conditions for a (strong) decoherence-free
subspace to be a stabilizer code.
In this section we are going to elaborate over these conditions in order to
connect (strong) decoherence-free stabilizer code, defined over a particular
type of errors, with classical code. For this purpose, this section is divided in three parts.
In the first part, we present some motivations for the tools constructed in the second part.
Then, in the second part, it is put forward a vector space
where the sum of vectors is related to the composition of operators.
Lastly, part three connects stabilizer codes to these vector spaces by means of an isomorphism.

\subsection{Motivation}

In order to illustrate the connection between DFS (sDFS) and
$\mathcal{S}_\text{DFS}$ ($\mathcal{S}_\text{sDFS}$), we present the example below.

\begin{example}
	Consider the same evolution as in Example~\ref{Example1}, but with
      \begin{equation}
            J = \frac{s+c}{2}\bigotimes_{j=1}^5 (\sigma_{xj} + i\sigma_{yj} + \sigma_{zj}),
      \end{equation}
      For $r\neq 0$, the eigenvectors of $J$ are all possible tensor products of $\ket{\psi_{+}}$ and $\ket{\psi_{-}}$, where
      $\ket{\psi_{+}} = \ket{0}$ and $\ket{\psi_{-}} = \frac{\ket{0} - \ket{1}}{\sqrt{2}}$.
      The eigenvalue of $\ket{\psi_{\pm}}$ is equal to  $\pm(\frac{s+c}{2})$.
      Thus, if the number of $\ket{\psi_{\pm}}$ in the eigenvector of $J$ is $n_{\pm}$, the eigenvalue of such
      eigenvector is $\frac{s+c}{2}$ if $n_{+} > n_{-}$ or $-\frac{s+c}{2}$
      if $n_{+} < n_{-}$. For $n_{+} > n_{-}$, the stabilizer set is given by
	\begin{equation}
		\mathcal{S}_\text{DFS} = \Big{\langle} \frac{2J}{s+c}\Big{\rangle} = \Big{\{}\Big{(}\frac{2J}{s+c}\Big{)}^i \colon i\in\mathbb{N}\Big{\}}.
		\label{Eq:Sdfs0}
	\end{equation}It is clear that $\mathcal{S}_\text{DFS}$ is abelian. Hence, we can construct a stabilizer code form $\mathcal{S}_\text{DFS}$.
	Additionally, from the commutation relation of Proposition~\ref{Prop:CommutationRelation} and
	$J = \frac{s+c}{2}\bigotimes_{j=1}^5\Big{(}\sigma_{xj} + i\sigma_{yj} + \sigma_{zj}\Big{)}$, the system Hamiltonian with the form
	\begin{equation}
		H_S = \bigotimes_{j=1}^5\eta_0\sigma_{yj},
		\label{Eq:Hs}
	\end{equation}where $\eta_0 = \frac{\gamma(s+c)^{2}}{4}$, belongs to $C_{G_N}(\mathcal{S}_\text{DFS})$. 
	Observe that if we have derived the form of $H_S$ in a similar way as the method in Example~\ref{Example1}; i.e., by
	imposing that the eigenvalue of $[H_S + \frac{i\gamma(s+c)}{4}(J-J^\dagger), J]$ over the
	subspace $\mathcal{Q}$ is equal to zero, we would obtain the same result as in the current example, which imposes
	the stronger condition $[H_S +  \frac{i\gamma(s+c)}{4}(J-J^\dagger), J] = 0$.
	This means that for the model of $N$
	qubits considered, the existence of a decoherence-free subspace
	is equivalent to the existence of a decoherence-free stabilizer code.
	\label{Example2}
\end{example}

There are several interesting aspects of dealing with stabilizer groups. One is the complexity reduction in defining
the stabilizer code, i.e., instead of using vector space basis we can use the group generator of the stabilizer group.
Encoding, error detection, error correction,
and decoding schemes can also be computationally-efficient constructed from the
stabilizer group. Another one is the connection between stabilizer groups and additive codes. This connection is implemented by means of an
isomorphism. Thus, statements over the equivalent additive code are directly translated to the stabilizer group and, more importantly, to the
stabilizer code. This approach is largely utilized to show existence or non-existence of stabilizer codes with specific parameters or properties.
In the following, we sketch the connection between stabilizer groups constructed from open quantum systems and additive codes.
In particular, the isomorphism utilized to connect them is introduced.

\begin{example}
	Following Example~\ref{Example2}, the stabilizer elements $S_1$ and $S_2$ in $\mathcal{S}_\text{DFS}$ can be written as
	\begin{eqnarray}
		S_1 &=& \bigotimes_{j = 1}^5\Big{(}\sigma_{xj} + i\sigma_{yj} + \sigma_{zj}\Big{)},\\
		S_2 &=& \mathbb{I}\otimes\mathbb{I}\otimes\mathbb{I}\otimes\mathbb{I}\otimes\mathbb{I}.
	\end{eqnarray}Now, we can create a map that represents errors as vectors. Let us consider
	\begin{eqnarray}
	\zeta\colon \tilde{G}_N &\rightarrow& \mathbb{C}^{4N},\nonumber\\
				\bigotimes_{j=1}^N \Big{(}a_{0j}\mathbb{I}_{j} + a_{1j}\sigma_{xj} + a_{2j}\sigma_{yj} + a_{3j}\sigma_{zj}\Big{)} &\mapsto& (a_{01}, \ldots, a_{0N}, a_{11}, \ldots, a_{1N}, a_{21}, \ldots, a_{2N}, a_{31}, \ldots a_{3N}),
	\label{Eq:Map1}
	\end{eqnarray}where $\tilde{G}_N\subseteq{G}_N$ has elements with the above description.
	As an example, applying the map $\zeta$ to the generators of $\mathcal{S}_\text{DFS}$ gives
	\begin{eqnarray}
		&\bm{v}_{S_1} = \zeta(S_1) =
		&\begin{pmatrix}
			0, & 0, & 0, & 0, & 0, & 1, & 1, & 1, & 1, & 1, & i, & i, & i, & i, & i, & 1, & 1, & 1, & 1, & 1
		\end{pmatrix}
	\end{eqnarray}and $\bm{v}_{S_2} = \zeta(S_2) = (1, 1, 1, 1, 1, 0, 0, 0, 0, 0, 0, 0, 0, 0, 0, 0, 0, 0, 0, 0)$.
	We can also examine the image of $H_S$, given in Eq.~(\ref{Eq:Hs}), by the map $\zeta$,
	\begin{eqnarray}
		&\bm{v} = \zeta(H_S) =
		&\begin{pmatrix}
			0, & 0, & 0, & 0, & 0, & 0, & 0, & 0, & 0, & 0, & 1, & 1, & 1, & 1, & 1, & 0, & 0, & 0, & 0, & 0
		\end{pmatrix}.
	\end{eqnarray}The commutativity property between $H_{ev}$ and $J$ can also be described using our representation.
	Let $\bm{v}_{H_{ev}} = \zeta(H_S + \frac{i\gamma}{2}\sqrt{sc}(J-J^\dagger))$ and write it as
	\begin{equation}
		\bm{v}_{H_{ev}} =
		\begin{pmatrix}
		\bm{b}_{0}, & \bm{b}_{1}, & \bm{b}_{2}, & \bm{b}_{3}
		\end{pmatrix},
	\end{equation}where $\bm{b}_{j}\in\mathbb{C}^{5}$, for all $j=0,1,2,3$. Similarly, $\bm{v}_{S_1}$ is given by
	\begin{equation}
		\bm{v}_{S_1} =
		\begin{pmatrix}
		\bm{a}_{0}, & \bm{a}_{1}, & \bm{a}_{2}, & \bm{a}_{3}
		\end{pmatrix},
	\end{equation}where $\bm{a}_{j}\in\mathbb{C}^{5}$, for all $j=0,1,2,3$.
	Then the commutation relation of Proposition~\ref{Prop:CommutationRelation} can be extended by defining
	\begin{eqnarray}
		\langle \bm{v}_{H_{ev}}, \bm{v}_{S_1}\rangle_{\zeta_{(1,j)}} &:=& (a_{2j} b_{3j} - a_{3j} b_{2j}),\\
		\langle \bm{v}_{H_{ev}}, \bm{v}_{S_1}\rangle_{\zeta_{(2,j)}} &:=& (a_{3j} b_{1j} - a_{1j} b_{3j}),\\
		\langle \bm{v}_{H_{ev}}, \bm{v}_{S_1}\rangle_{\zeta_{(3,j)}} &:=& (a_{1j} b_{2j} - a_{2j} b_{1j}),
		\label{Eq:DualSpace}
	\end{eqnarray}for $j = 1,\ldots, N$. We can see that $[H_{ev}, J] = 0$ if and only if
	$\langle \bm{v}_{H_{ev}}, \bm{v}_{S_1}\rangle_{\zeta_{(1,j)}} = \langle \bm{v}_{H_{ev}}, \bm{v}_{S_1}\rangle_{\zeta_{(2,j)}} = \langle \bm{v}_{H_{ev}}, \bm{v}_{S_1}\rangle_{\zeta_{(3,j)}} = 0$,
	for $j=1,\ldots,N$. In our example,
	we have $\bm{v}_{H_{ev}} = (0, 0, \ldots, 0)\in\mathbb{C}^{20}$, which gives
	$\langle \bm{v}_{H_{ev}}, \bm{v}_{S_1}\rangle_{\zeta_{(1,j)}} = \langle \bm{v}_{H_{ev}}, \bm{v}_{S_1}\rangle_{\zeta_{(2,j)}} = \langle \bm{v}_{H_{ev}}, \bm{v}_{S_1}\rangle_{\zeta_{(3,j)}} = 0$,
	for $j = 1,\ldots, N$.

	Consider the composition of operators. Let $E_1,E_2\in G^N$ be two errors written as
	\begin{eqnarray}
		E_1 &=& \bigotimes_{j=1}^N \Big{(}a_{0j}\mathbb{I}_{j} + a_{1j}\sigma_{xj} + a_{2j}\sigma_{yj} + a_{3j}\sigma_{zj}\Big{)},\\
		E_2 &=& \bigotimes_{j=1}^N \Big{(}b_{0j}\mathbb{I}_{j} + b_{1j}\sigma_{xj} + b_{2j}\sigma_{yj} + b_{3j}\sigma_{zj}\Big{)}.
	\end{eqnarray}Then,
	\begin{equation}
		E_1 E_2 = \bigotimes_{j=1}^N\Big{(}c_{0j}\mathbb{I}_{j} + c_{1j}\sigma_{xj} + c_{2j}\sigma_{yj} + c_{3j}\sigma_{zj}\Big{)},
	\end{equation}where
	\begin{subequations}
		\begin{eqnarray}
			c_{0j} &=& a_{0j} b_{0j} + a_{1j} b_{1j}  +   a_{2j} b_{2j} + a_{3j} b_{3j},\\
			c_{1j} &=& (a_{1j} b_{0j} + a_{0j} b_{1j}) + i(a_{2j} b_{3j} - a_{3j} b_{2j}),\\
			c_{2j} &=& (a_{2j} b_{0j} + a_{0j} b_{2j}) + i(a_{3j} b_{1j} - a_{1j} b_{3j}),\\
			c_{3j} &=& (a_{3j} b_{0j} + a_{0j} b_{3j}) + i(a_{1j} b_{2j} - a_{2j} b_{1j}),
		\end{eqnarray}
		\label{Eq:c_ij}
	\end{subequations}for $j=1,\ldots,N$. Thus, we need to impose over the map $\zeta$ the following condition for composition of operators
	\begin{eqnarray}
		\zeta(E_1 E_2) =
				\begin{pmatrix}
					c_{01}, & \ldots, & c_{0N}, & c_{11}, & \ldots, & c_{1N}, & c_{21}, & \ldots, & c_{2N}, & c_{31}, &\ldots, & c_{3N}
				\end{pmatrix},
	\label{Eq:HomomorphismGN}
	\end{eqnarray}where $c_{ij}$, for $i\in\{0,1,2,3\}$ and $j\in\{1,\ldots,N\}$, is defined in Eq.~(\ref{Eq:c_ij}).
	Further details of this operation will be given below and in the following subsections.
	\label{Example3}
\end{example}

Let us introduce the map $\zeta$ in formal terms and derive an additive operation from this definition.

\begin{definition}
	Let $E_1,E_2\in G^N$ be two errors written as
	\begin{eqnarray}
		E_1 &=& \bigotimes_{j=1}^N \Big{(}a_{0j}\mathbb{I}_{j} + a_{1j}\sigma_{xj} + a_{2j}\sigma_{yj} + a_{3j}\sigma_{zj}\Big{)},\\
		E_2 &=& \bigotimes_{j=1}^N \Big{(}b_{0j}\mathbb{I}_{j} + b_{1j}\sigma_{xj} + b_{2j}\sigma_{yj} + b_{3j}\sigma_{zj}\Big{)}.
	\end{eqnarray}Then we define the map
	\begin{eqnarray}
		\zeta\colon \tilde{G}_N &\rightarrow& \mathbb{C}^{4N},\nonumber\\
		\bigotimes_{j=1}^N \Big{(}a_{0j}\mathbb{I}_{j} + a_{1j}\sigma_{xj} + a_{2j}\sigma_{yj} + a_{3j}\sigma_{zj}\Big{)} &\mapsto& (a_{01}, \ldots, a_{0N}, a_{11}, \ldots, a_{1N}, a_{21}, \ldots, a_{2N}, a_{31}, \ldots a_{3N})
	\end{eqnarray}via the operation
	\begin{eqnarray}
		\zeta(E_1 E_2) =
				\begin{pmatrix}
					c_{01}, & \ldots, & c_{0N}, & c_{11}, & \ldots, & c_{1N}, & c_{21}, & \ldots, & c_{2N}, & c_{31}, & \ldots, & c_{3N}\\
				\end{pmatrix},
	\label{Eq:HomomorphismGN}
	\end{eqnarray}where
	\begin{subequations}
		\begin{eqnarray}
			c_{0j} &=& a_{0j} b_{0j} + a_{1j} b_{1j} + a_{2j} b_{2j} + a_{3j} b_{3j},\\
			c_{1j} &=& (a_{1j} b_{0j} + a_{0j} b_{1j}) + i(a_{2j} b_{3j} - a_{3j} b_{2j}),\\
			c_{2j} &=& (a_{2j} b_{0j} + a_{0j} b_{2j}) + i(a_{3j} b_{1j} - a_{1j} b_{3j}),\\
			c_{3j} &=& (a_{3j} b_{0j} + a_{0j} b_{3j}) + i(a_{1j} b_{2j} - a_{2j} b_{1j}),
		\end{eqnarray}
	\end{subequations}for $j=1,\ldots,N$.
	\label{Def:HomomorphismGN}
\end{definition}

\begin{definition}
\label{Def:zetaSum}
	Let $N$ be a positive integer, and $\bm{v}_1,\bm{v}_2\in\mathbb{C}^{4N}$ be two vectors given, respectively, by
	\begin{eqnarray}
		\bm{v}_1 &=&
		\begin{pmatrix}
			a_{01}, & \ldots, & a_{0N}, & a_{11}, & \ldots, & a_{1N}, & a_{21}, & \ldots, & a_{2N}, & a_{31}, & \ldots, & a_{3N}
		\end{pmatrix},\\
		\bm{v}_2 &=&
		\begin{pmatrix}
			b_{01}, & \ldots, & b_{0N}, & b_{11}, & \ldots, & b_{1N}, & b_{21}, & \ldots, & b_{2N}, & b_{31}, & \ldots, & b_{3N}
		\end{pmatrix}.
	\end{eqnarray}Define the binary operation $+_{\zeta}$ as
	\begin{equation}
		\bm{v}_1 +_{\zeta} \bm{v}_2 :=
		\begin{pmatrix}
			c_{01}, & \ldots, & c_{0N}, & c_{11}, & \ldots, & c_{1N}, & c_{21}, & \ldots, & c_{2N}, & c_{31}, & \ldots, & c_{3N}
		\end{pmatrix},
	\end{equation}where
	\begin{subequations}
		\begin{eqnarray}
			c_{0j} &=& a_{0j} b_{0j} + a_{1j} b_{1j} + a_{2j} b_{2j} + a_{3j} b_{3j},\\
			c_{1j} &=& (a_{1j} b_{0j} + a_{0j} b_{1j}) + i(a_{2j} b_{3j} - a_{3j} b_{2j}),\\
			c_{2j} &=& (a_{2j} b_{0j} + a_{0j} b_{2j}) + i(a_{3j} b_{1j} - a_{1j} b_{3j}),\\
			c_{3j} &=& (a_{3j} b_{0j} + a_{0j} b_{3j}) + i(a_{1j} b_{2j} - a_{2j} b_{1j}),
		\end{eqnarray}
		\label{Eq:Mult}
	\end{subequations}for $j=1,\ldots,N$.
\end{definition}

Considering the $+_\zeta$ operation defined above as the sum operation of the additive codes, we derive some constraint over the coordinates of
the elements in these codes.

\begin{proposition}
	Let $C$ be an $+_\zeta$-additive code. If $\bm{v}_1 = (\bm{a}_0,\bm{a}_1,\bm{a}_2,\bm{a}_3)$ and
	$\bm{v}_2 = (\bm{b}_0,\bm{b}_1,\bm{b}_2,\bm{b}_3)$ are elements in $C$, then
	\begin{subequations}
		\begin{eqnarray}
			a_{2j} b_{3j} &=& a_{3j} b_{2j},\\
			a_{3j} b_{1j} &=& a_{1j} b_{3j},\\
			a_{1j} b_{2j} &=& a_{2j} b_{1j},
		\end{eqnarray}
	\label{condition_one}
	\end{subequations}and the following system of equations must also be satisfied
	\begin{subequations}
		\begin{eqnarray}
			a_{lj} &=& 0,\\
			a_{ij} &=& \pm ia_{kj},
		\end{eqnarray}
	\label{condition_two}
	\end{subequations}for pairwise distinct $l,i,k\in\{1,2,3\}$ and each $j=1,\ldots,N$.
\end{proposition}
\begin{proof}
	The set of conditions presented in Eq.~(\ref{condition_one}) follows by imposing commutativity of $\bm{v}_1 +_\zeta \bm{v}_2$
	and $\bm{v}_2 +_\zeta \bm{v}_1$ in Eq.~(\ref{Eq:Mult}). To derive the conditions in Eq.~(\ref{condition_two}), notice that
	Eq.~(\ref{condition_one}) can be described as
	\begin{subequations}
		\begin{eqnarray}
			a_{2j} b_{3j} - a_{3j} b_{2j} &=& 0,\\
			a_{3j} b_{1j} - a_{1j} b_{3j} &=& 0,\\
			a_{1j} b_{2j} - a_{2j} b_{1j} &=& 0,
		\end{eqnarray}
	\label{system_one}
	\end{subequations}which has nontrivial solution if and only if $a_{1j} a_{2j} a_{3j} = 0$. Substituting this condition
	in Eq.~(\ref{system_one}) and imposing nontriviality to the solution again, we obtain
	$a_{ij}^2 = - a_{kj}^2$ and $a_{lj} = 0$ for pairwise distinct $l,i,k\in\{1,2,3\}$. Notice that for each $j$,
	we have independent conditions.
\end{proof}

We have presented some intuitions on how to relate operators and vectors. Some constraints on the coordinates of the
vectors have been presented. However, we need to develop further tools and properties to derive a stabilizer formalism
connecting stabilizer code and additive codes. In particular, three points are covered in the following subsection.
Firstly, we demonstrate that the map $\langle \cdot, \cdot \rangle_\zeta$ is a symplectic form.
Using this fact, we show that the map $\zeta$ is an isomorphism between abelian sets of operators and additive codes.
Lastly, we introduce symplectic dual codes and the stabilizer formalism connecting quantum stabilizer codes with
$+_\zeta$-additive codes.

\subsection{Symplectic form and Additive Codes}
\label{SubSec:VectorSpace}

A symplectic form connects the centralizer of a stabilizer group to the dual code of the classical code corresponding
to the stabilizer group. Symplectic forms can be defined over vector spaces or groups. In the following we consider
a symplectic form over groups. Thus, the dual code obtained is an additive code.

\begin{definition}
	A symplectic form over an additive group $\mathcal{G}$ to a field $F$ is a function
	\begin{eqnarray}
		f\colon \mathcal{G}\times\mathcal{G}&\rightarrow& F\\
		(g_1,g_2)&\mapsto& f(g_1,g_2),
	\end{eqnarray}such that
	\begin{subequations}
		\begin{eqnarray}
			f(g_1+g_2,g_3) &=& f(g_1,g_3) + f(g_2,g_3),\\
			f(g_1,g_2)     &=& -f(g_2,g_1),\\
			f(g_1,g_1)     &=& 0,
		\end{eqnarray}
	\end{subequations}for all $g_1,g_2,g_3\in\mathcal{G}$.
	\label{Def:SympForm}
\end{definition}

For the operation in Eq.~(\ref{Eq:DualSpace}) to be a symplectic form, the first point we need to show is that the image of
$\zeta$ equipped with a proper additive operation forms an additive group.

We claim that the set $\mathcal{V} = \zeta(C_{G_N}(\mathcal{S}))$, where $\mathcal{S}$ is a stabilizer group, equipped with $+_\zeta$
operation from Definition~\ref{Def:zetaSum} is an additive group. Indeed, let $\bm{v}_A,\bm{v}_B,\bm{v}_C\in\mathcal{V}$,
then the following axioms are satisfied:
\begin{enumerate}
	\item $\mathcal{V}$ is closed under $+_\zeta$;
	\item $\bm{v}_A+_\zeta \bm{v}_B = \bm{v}_B +_\zeta \bm{v}_A$;
	\item $(\bm{v}_A+_\zeta \bm{v}_B)+_\zeta \bm{v}_C = \bm{v}_A+_\zeta (\bm{v}_B+_\zeta \bm{v}_C)$;
	\item there exists an element $\bm{v}_{\mathbb{I}}$ such that $\bm{v}_A +_\zeta \bm{v}_{\mathbb{I}} = \bm{v}_A$;
	\item For each $\bm{v}_A\in\mathcal{V}$, there exists an element $\bm{v}_B\in\mathcal{V}$ such that
		  $\bm{v}_A +_\zeta \bm{v}_B = \bm{v}_{\mathbb{I}} = \bm{v}_B +_\zeta \bm{v}_A$.
\end{enumerate}The first point is clearly true. For the second point, we have that $\mathcal{V}$ is the image of
$\zeta$ over $C_{G_N}(\mathcal{S})$. From Proposition~\ref{Prop:CommutationRelation}, we have
\begin{subequations}
	\begin{eqnarray}
		a_{2j} b_{3j} - a_{3j} b_{2j} = 0,\\
		a_{3j} b_{1j} - a_{1j} b_{3j} = 0,\\
		a_{1j} b_{2j} - a_{2j} b_{1j} = 0,
	\end{eqnarray}
	\label{Eq:Comm}
\end{subequations}for $j=1,\ldots,N$, where $a_{lj}$ and $b_{pj}$ are the coordinates of the
vectors $\bm{v}_A$ and $\bm{v}_B$, respectively, for $l,p=1,2,3$. Thus, we can see from Definition~\ref{Def:zetaSum}
that $+_\zeta$ is abelian. For the third point, let $\bm{v}_D = \bm{v}_A +_\zeta \bm{v}_B$ and
$\bm{v}_E = \bm{v}_B +_\zeta \bm{v}_C$, where each coordinate is given by
\begin{subequations}
	\begin{eqnarray}
		d_{0j} = a_{0j} b_{0j} + a_{1j} b_{1j} + a_{2j} b_{2j} + a_{3j} b_{3j},\\
		d_{1j} = a_{1j} b_{0j} + a_{0j} b_{1j},\\
		d_{2j} = a_{2j} b_{0j} + a_{0j} b_{2j},\\
		d_{3j} = a_{3j} b_{0j} + a_{0j} b_{3j},
	\end{eqnarray}
\end{subequations}and
\begin{subequations}
	\begin{eqnarray}
		e_{0j} = b_{0j} c_{0j} + b_{1j} c_{1j} + b_{2j} c_{2j} + b_{3j} c_{3j},\\
		e_{1j} = b_{1j} c_{0j} + b_{0j} c_{1j},\\
		e_{2j} = b_{2j} c_{0j} + b_{0j} c_{2j},\\
		e_{3j} = b_{3j} c_{0j} + b_{0j} c_{3j},
	\end{eqnarray}
\end{subequations}for $j=1,\ldots,N$. Then, the result of the sum $\bm{v}_F = \bm{v}_D + \bm{v}_C$ can be described by
\begin{eqnarray}
	f_{0j} &=& (a_{0j} b_{0j} + a_{1j} b_{1j} + a_{2j} b_{2j} + a_{3j} b_{3j}) c_{0j} + (a_{1j} b_{0j} + a_{0j} b_{1j}) c_{1j}\nonumber\\
	 	   &+& (a_{2j} b_{0j} + a_{0j} b_{2j}) c_{2j} + (a_{3j} b_{0j} + a_{0j} b_{3j}) c_{3j},\nonumber\\
	f_{1j} &=& (a_{1j} b_{0j} + a_{0j} b_{1j}) c_{0j} + (a_{0j} b_{0j} + a_{1j} b_{1j} + a_{2j} b_{2j} + a_{3j} b_{3j}) c_{1j},\nonumber\\
	f_{2j} &=& (a_{2j} b_{0j} + a_{0j} b_{2j}) c_{0j} + (a_{0j} b_{0j} + a_{1j} b_{1j} + a_{2j} b_{2j} + a_{3j} b_{3j}) c_{2j},\nonumber\\
	f_{3j} &=& (a_{3j} b_{0j} + a_{0j} b_{3j}) c_{0j} + (a_{0j} b_{0j} + a_{1j} b_{1j} + a_{2j} b_{2j} + a_{3j} b_{3j}) c_{3j}.
\end{eqnarray}Similarly, it follows that the sum $\bm{v}_{F'} = \bm{v}_A + \bm{v}_E$ is equal to
\begin{eqnarray}
	f'_{0j} &=& a_{0j} (b_{0j} c_{0j} + b_{1j} c_{1j} + b_{2j} c_{2j} + b_{3j} c_{3j}) + a_{1j} (b_{1j} c_{0j} + b_{0j} c_{1j})\nonumber\\
		    &+& a_{2j} (b_{2j} c_{0j} + b_{0j} c_{2j}) + a_{3j} (b_{3j} c_{0j} + b_{0j} c_{3j}),\nonumber\\
	f'_{1j} &=& a_{1j} (b_{0j} c_{0j} + b_{1j} c_{1j} + b_{2j} c_{2j} + b_{3j} c_{3j}) + a_{0j} (b_{1j} c_{0j} + b_{0j} c_{1j}),\nonumber\\
	f'_{2j} &=& a_{2j} (b_{0j} c_{0j} + b_{1j} c_{1j} + b_{2j} c_{2j} + b_{3j} c_{3j}) + a_{0j} (b_{2j} c_{0j} + b_{0j} c_{2j}),\nonumber\\
	f'_{3j} &=& a_{3j} (b_{0j} c_{0j} + b_{1j} c_{1j} + b_{2j} c_{2j} + b_{3j} c_{3j}) + a_{0j} (b_{3j} c_{0j} + b_{0j} c_{3j}).
\label{Eq:fprime}
\end{eqnarray}Rearranging the terms in Eq.~(\ref{Eq:fprime}) and utilizing the relation from Eq.~(\ref{Eq:Comm}), we see that
$f_{ij} = f'_{ij}$ for $i=0,1,2,3$ and $j=1,\ldots, N$. Therefore, we have proven Property 3. From the definition of $+_\zeta$ and
the relation from Eq.~(\ref{Eq:Comm}), we have that the identity element exists. In particular, the identity
element is given by $\bm{v}_{\mathbb{I}} = (\bm{1}_{N},\bm{0}_{N},\bm{0}_{N},\bm{0}_{N})$, where $\bm{1}_{N}$ and $\bm{0}_{N}$ are
$N$-dimensional vectors with all coordinates equal to $1$ and $0$, respectively. The same approach can be used to show Property 5.

Now, we can use the previous algebraic structure to show that the expression given in Eq.~(\ref{Eq:DualSpace})
is a symplectic form. 

\begin{proposition}
Let $N$ be a positive integer and
$\mathcal{V} = \{\bm{v}\in\mathbb{C}^{4N}|\bm{v} = (\bm{x}_0, \bm{x}_1, \bm{x}_2, \bm{x}_3)\text{ where }\bm{x}_0 = (1, 1,\ldots,1)\in\mathbb{C}^N\text{ and }\bm{x}_1,\bm{x}_2,\bm{x}_3\in\mathbb{C}^N\}$
be a group under $+_\zeta$. Then the maps

\begin{eqnarray}
	\langle\cdot, \cdot\rangle_{\zeta_{(1,j)}}\colon \mathbb{C}^{4N}\times\mathbb{C}^{4N}&\rightarrow& \mathbb{C}\nonumber\\
	(\bm{v}_A,\bm{v}_B)&\mapsto& \langle \bm{v}_A,\bm{v}_B\rangle_{\zeta_{(1,j)}} = (a_{2j} b_{3j} - a_{3j} b_{2j}),
	\label{Eq:Symp1}
\end{eqnarray}
\begin{eqnarray}
	\langle\cdot, \cdot\rangle_{\zeta_{(2,j)}}\colon \mathbb{C}^{4N}\times\mathbb{C}^{4N}&\rightarrow& \mathbb{C}\nonumber\\
	(\bm{v}_A,\bm{v}_B)&\mapsto&\langle \bm{v}_A,\bm{v}_B\rangle_{\zeta_{(2,j)}} = (a_{3j} b_{1j} - a_{1j} b_{3j}),
	\label{Eq:Symp2}
\end{eqnarray}
\begin{eqnarray}
	\langle\cdot, \cdot\rangle_{\zeta_{(3,j)}}\colon \mathbb{C}^{4N}\times\mathbb{C}^{4N}&\rightarrow& \mathbb{C}\nonumber\\
	(\bm{v}_A,\bm{v}_B)&\mapsto&\langle \bm{v}_A,\bm{v}_B\rangle_{\zeta_{(3,j)}} = (a_{1j} b_{2j} - a_{2j} b_{1j}),
	\label{Eq:Symp3}
\end{eqnarray}are symplectic forms over $\mathcal{V}$, where $\bm{v}_A = (\bm{x}_0, \bm{a}_1, \bm{a}_2, \bm{a}_3)$,
$\bm{v}_B = (\bm{x}_0, \bm{b}_1, \bm{b}_2, \bm{b}_3)$, and $j=1,\ldots,N$.
\end{proposition}

\begin{proof}
	Let $\bm{v}_A = (\bm{x}_0, \bm{a}_1, \bm{a}_2, \bm{a}_3), \bm{v}_B = (\bm{x}_0, \bm{b}_1, \bm{b}_2, \bm{b}_3),$
	and $\bm{v}_C = (\bm{x}_0, \bm{c}_1, \bm{c}_2, \bm{c}_3)\in\mathcal{V}$.
	From the clear relation between Eqs.~(\ref{Eq:Symp1}-\ref{Eq:Symp3}), we only need to show that one
	of these functions is a symplectic form.	Then,
	\begin{eqnarray}
		\langle \bm{v}_A +_\zeta \bm{v}_B, \bm{v}_C\rangle_{\zeta_{(1,j)}} &=& (a_{2j}x_{0j} + x_{0j}b_{2j}) c_{3j} - (a_{3j}x_{0j} + a_{0j}x_{3j}) c_{2j}\nonumber\\
													 &=& (a_{2j} c_{3j} - a_{3j}c_{2j})x_{0j}  + (b_{2j} c_{3j} - b_{3j} c_{2j}) x_{0j}\nonumber\\
													 &=& \langle \bm{v}_A , \bm{v}_C\rangle_{\zeta_{(1,j)}}  + \langle \bm{v}_B, \bm{v}_C\rangle_{\zeta_{(1,j)}},
	\end{eqnarray}where $j=1,\ldots,N$ and we have used the fact that
	$\bm{x}_{0} = (1,1,\ldots,1)$. It is also possible to see that
	\begin{eqnarray}
		\langle \bm{v}_A, \bm{v}_B \rangle_{\zeta_{(1,j)}} &=& a_{2j} b_{3j} - a_{3j}b_{2j}\nonumber\\
										   &=& - (a_{3j}b_{2j} - a_{2j} b_{3j})\nonumber\\
										   &=& - \langle \bm{v}_B, \bm{v}_A \rangle_{\zeta_{(1,j)}},
	\end{eqnarray}and $\langle \bm{v}_A, \bm{v}_A \rangle_{\zeta_{(1,j)}} = 0$.
	Thus, we have shown that $\langle\cdot, \cdot\rangle_{\zeta_{(1,j)}}$ is, in fact, a symplectic form.
\end{proof}

\begin{example}
	Let $G_3$ be the error set for three qubit systems and $A,B\in G_3$ be operators given by
	\begin{eqnarray}
		A &=& (\mathbb{I} + i\sigma_y + \sigma_z)\otimes(\mathbb{I} + \sigma_x - i\sigma_y)\otimes\mathbb{I}\\
		B &=& \mathbb{I}\otimes\mathbb{I}\otimes(\mathbb{I} + \sigma_x - i\sigma_z).
	\end{eqnarray}Then, on the one hand, we have that
	\begin{equation}
		A  B = (\mathbb{I} + i\sigma_y + \sigma_z)\otimes(\mathbb{I} + \sigma_x - i\sigma_y)\otimes(\mathbb{I} + \sigma_x - i\sigma_z),
	\end{equation}which implies
	\begin{eqnarray}
		\zeta(A B) = (1, 1, 1,     0, 1, 1,      i, -i, 0,       1, 0, -i).
		\label{Eq:ExampleEasy1}
	\end{eqnarray}On the other hand, the action of $\zeta$ on $A$ and $B$ is given by
	\begin{eqnarray}
		\bm{v}_A &=& \zeta(A) = (1,1,1, 0,1,0, i,-i,0, 1,0,0)\\
		\bm{v}_B &=& \zeta(B) = (1,1,1, 0,0,1, 0,0,0, 0,0,-i).
	\end{eqnarray}Therefore,
	\begin{eqnarray}
		\bm{v}_A +_\zeta \bm{v}_B = (1, 1, 1,     0, 1, 1,      i, -i, 0,       1, 0, -i),
		\label{Eq:ExampleEasy2}
	\end{eqnarray}where we used Definition~\ref{Def:zetaSum}. From Eqs.~(\ref{Eq:ExampleEasy1})~and~(\ref{Eq:ExampleEasy2})
	we see that
	\begin{equation}
		\zeta(A B) = \bm{v}_A +_\zeta \bm{v}_B.
	\end{equation}
\end{example}

Now, we have the tools to define the symplectic dual of a $+_\zeta$-additive code.

\begin{definition}
	Let $N$ be a positive integer and
	$C = \{\bm{c}\in\mathbb{C}^{4N}|\bm{c} = (\bm{c}_0, \bm{c}_1, \bm{c}_2, \bm{c}_3),\text{ where }\bm{c}_0 = (1, 1,\ldots,1)\in\mathbb{C}^N\text{ and }\bm{c}_1,\bm{c}_2,\bm{c}_3\in\mathbb{C}^N\}$ be an $+_\zeta$-additive code.
	The symplectic dual of $C$ is given by
	\begin{equation}
		C^{\perp_\zeta} := \{\bm{c}\in\mathbb{C}^{4N}\colon \langle \bm{c}, \bm{d}\rangle_{\zeta_{(l,j)}} = 0, \text{ for all }\bm{d}\in C, l=1,2,3, \text{ and }j=1,\ldots,N\}.
	\end{equation}
\end{definition}

Similar to previous works on stabilizer codes, we are going to derive a connection between stabilizer codes
and classical error-correcting codes. This approach enables us to derive algebraic conditions for the construction
and existence of decoherence-free stabilizer codes. We can use it to show nonexistence of decoherence-free stabilizer codes with some specific parameters.

\begin{theorem}
	Let $\mathcal{V}_{\mathcal{S}_\text{DFS}} = \zeta(\mathcal{S}_\text{DFS})$ or
	$\mathcal{V}_{\mathcal{S}_\text{sDFS}} = \zeta(\mathcal{S}_\text{sDFS})$ be a basis of the
	$+_\zeta$-additive code of the form
	$C = \{\bm{c}\in\mathbb{C}^{4N}|\bm{c} = (\bm{c}_0, \bm{c}_1, \bm{c}_2, \bm{c}_3)\text{ where }\bm{c}_0 = (1, 1,\ldots,1)\in\mathbb{C}^N\text{ and }\bm{c}_1,\bm{c}_2,\bm{c}_3\in\mathbb{C}^N\}$. Then,
		\begin{enumerate}
			\item A decoherence-free stabilizer code $\mathcal{Q}$ exists if there exists an $+_\zeta$-additive code $C$ over
			$\mathbb{C}$ generated by $\mathcal{V}_{\mathcal{S}_\text{DFS}}$ such that $C\leq C^{\perp_\zeta}$ and
			$\zeta(H_{ev})\in C^{\perp_\zeta}$;
			\item A strong decoherence-free stabilizer code $\mathcal{Q}$ exists if there exists an $+_\zeta$-additive code
			$C$ over $\mathbb{C}$ generated by $\mathcal{V}_{\mathcal{S}_\text{sDFS}}$ such that $C\leq C^{\perp_\zeta}$ and
			$\zeta(H_{S})\in C^{\perp_\zeta}$.
		\end{enumerate}
	\label{Thm:StabilizerFormParticular}
\end{theorem}

\begin{proof}
	First of all, since $C\leq C^{\perp_\zeta}$, then for all $S_1,S_2\in\mathcal{S}_\text{DFS}$ we have $[S_1,S_2] = 0$. This implies the existence of a
	maximum jointly eigenspace of all operators in $\mathcal{S}_\text{DFS}$. Let us denote it by $\mathcal{Q}$. In particular, $\mathcal{Q}$ is a stabilizer code
	with stabilizer given by $\mathcal{S}_\text{DFS}$. On the other hand, the hypothesis $\zeta(H_{ev})\in C^{\perp_\zeta}$ leads to
	$[H_{ev},S_i]=0$ for any $S_i\in\mathcal{S}_\text{DFS}$. Therefore, the eigenvalue of the commutator of $H_{ev}$ with any operator
	in $\mathcal{S}_\text{DFS}$ is equal to zero. Using Proposition~\ref{Proposition:DFS}, we have that $\mathcal{Q}$ is also a decoherence-free subspace.

	The same strategy can be used to deduce the second claim.
\end{proof}

\section{Decoherence-Free Stabilizer Codes for General Noise}
\label{sec:Main}
This section extends the previous results to general error operators. The approach followed to connect operators and classical codes
is based on matrix vectorization. After showing the corresponding vectorized operations, we demonstrate that the
formulation of Section~\ref{sec:Particular_main} is indeed a particular case of the current formulation. Additionally, the standard stabilizer
formalism can also be derived from our formalism.

\subsection{Motivation}

Suppose we wish to extend the formulation of the previous section to operators of the form
\begin{equation}
	E = \sum_{l=1}^L \bigotimes_{j=1}^N (a_{0j}^l\mathbb{I}_{j}+ a_{1j}^l\sigma_{xj} + a_{2j}^l\sigma_{yj} + a_{3j}^l\sigma_{zj}),
\end{equation}where $L$ is the number of terms in the sum describing the operator $E$, and $N$
is the number of physical systems. A naïve approach would be to map operators to matrices. There are some problems with this strategy.
First of all, one should impose an ordering over the terms in the sum going from $l=1,\ldots,L$ as a means to make a uniquely correspondence
between each term in the sum and a row in the matrix. Secondly, the composition of errors could result in a
sum of matrices giving a matrix with more rows than the original matrices that are being summed; e.g., suppose we have $E_1$ with
$L_1$ terms in the sum and $E_2$ with $L_2$ terms in the sum, then $E_1\circ E_2$ can produce up to $L_1\times L_2$ terms.
This can be solved since there is a maximum $L'$ of terms with which any operator can be described. Third, and more importantly,
the above representation is not unique. To see this, consider the operator
\begin{equation}
	A = (\mathbb{I} + \sigma_x + \sigma_z)\otimes\sigma_y + \sigma_z\otimes\sigma_x,
\end{equation}which can also be written as
\begin{equation}
	A = (\mathbb{I}+\sigma_x)\otimes\sigma_y + \sigma_z\otimes(\sigma_x + \sigma_y).
\end{equation}The issue of uniqueness in representing an operator and, consequently, its matrix representation
may be solved by introducing equivalence classes over matrix spaces similar to the equivalence classes utilized in
the definition of tensor product of vector spaces~\cite{Lang2005}.
Even tough these problem may be solved, the formulation seems not straightforward.
Therefore, in the following we use matrix vectorization to avoid all these complications.

Let $\{\ket{i}\}_{i=1}^q$ be a basis of a Hilbert space $\mathcal{H}_q$, and
$\ketbra{i}{j}\in\mathcal{L}(\mathcal{H}_q)$ be a linear operator over the Hilbert space $\mathcal{H}_q$.
Vectorization is a bijective linear map from $\mathcal{L}(\mathcal{H}_q)$ to $\mathcal{H}_q^{\otimes 2}$ defined as~\cite{Abadir2005}
\begin{equation}
	\text{vec}(\ketbra{i}{j}) := \ket{i}\ket{j}.
\end{equation}Vectorization can be extended to any operator space.
Let $\ketbra{i_1}{j_1}\otimes\ketbra{i_2}{j_2}\otimes\cdots\otimes\ketbra{i_N}{j_N}\in\mathcal{L}(\mathcal{H}_q^{\otimes N})$,
then
\begin{equation}
	\text{vec}(\ketbra{i_1}{j_1}\otimes\ketbra{i_2}{j_2}\otimes\cdots\otimes\ketbra{i_N}{j_N}) =
	\ket{i_1}\otimes\ket{i_2}\otimes\cdots\otimes\ket{i_N}\otimes\ket{j_1}\otimes\ket{j_2}\otimes\cdots\otimes\ket{j_N}.
\end{equation}Since $\ketbra{i_1}{j_1}\otimes\ketbra{i_2}{j_2}\otimes\cdots\otimes\ketbra{i_N}{j_N}$ forms a basis for the
space and the vectorization is a bijective linear map, it can be applied to any operator in
$\mathcal{L}(\mathcal{H}_q^{\otimes N})$.

Several properties can be derived for matrix vectorization. Two operations we have used are composition and commutation of operators. For the first, we can use the relation
\begin{equation}
	\text{vec}(ABC) = (A\otimes C^T)\text{vec}(B).
\end{equation}In particular, we have $\text{vec}(AB) = (A\otimes\mathbb{I})\text{vec}(B)$. The commutator can be
easily obtained from the above relation and the linearity of the vectorization. We have
\begin{equation}
	\text{vec}([A,B]) = (A\otimes\mathbb{I} - \mathbb{I}\otimes A^T)\text{vec}(B).
	\label{Eq:CommutatorVec}
\end{equation}

\begin{definition}
	Let $A,B\in\mathcal{L}(\mathcal{H}_q^N)$ be operators. We define the sum of the
	vectors $\text{vec}(A)$ and $\text{vec}(B)$ by
	\begin{equation}
		\text{vec}(A) +_{\text{vec}} \text{vec}(B) = (A\otimes\mathbb{I})\text{vec}(B).
	\end{equation}If $A$ commutes with $B$, then it is clear that
$\text{vec}(A) +_{\text{vec}} \text{vec}(B) = \text{vec}(B) +_{\text{vec}} \text{vec}(A)$.
Note that $+_{\text{vec}}$ is not the traditional sum of vectors, which always commutes.
\end{definition}

We utilize this relation to show that the result from the previous section and the standard
stabilizer formalism can be derived from the formulation presented below. Furthermore, the vectorization of
the commutator between two operators is used later to construct the symplectic form and the dual code of the
additive code.

\begin{proposition}
	Let $\mathcal{S}$ be a stabilizer set with operators satisfying the structure of the previous section. Assume that
	$\mathcal{C}_\zeta = \zeta(\mathcal{S})$ and $\mathcal{C}_\text{vec} = \text{vec}(\mathcal{S})$, where the composition
	of operators in $\mathcal{S}$ corresponds to the respective operation of the additive group. Then
	$\mathcal{C}_\zeta \equiv \mathcal{C}_\text{vec}$.
\end{proposition}

\begin{proof}
	First of all, consider a quantum system with $N = 1$. An operator $E$ can be written as
	$E = e_{01}\mathbb{I} + e_{11}\sigma_x + e_{21}\sigma_y + e_{31}\sigma_z$ or
	$E = e_{00}^1\ketbra{0}{0} + e_{01}^1\ketbra{0}{1} + e_{10}^1\ketbra{1}{0} + e_{11}^1\ketbra{1}{1}$, where
	$e_{i1}$, $i=0,1,2,3$, and $e_{pq}^1$, $p,q=0,1$, satisfy the relations
	\begin{subequations}
		\begin{eqnarray}
			e_{00}^1 &=& e_{01} + e_{31},\\
			e_{01}^1 &=& e_{11} - ie_{21},\\
			e_{10}^1 &=& e_{11} + ie_{21},\\
			e_{11}^1 &=& e_{01} - e_{31},
		\end{eqnarray}
	\end{subequations}and
	\begin{subequations}
		\begin{eqnarray}
			e_{01} &=& (e_{00}^1 + e_{11}^1)/2,\\
			e_{11} &=& (e_{01}^1 + e_{10}^1)/2,\\
			e_{21} &=& (e_{10}^1 - e_{01}^1)/2i,\\
			e_{31} &=& (e_{00}^1 - e_{11}^1)/2.
		\end{eqnarray}
	\end{subequations}Extending these relations to any positive integer $N$, taking into account that the relations
	are independent from one to another qubit, we obtain
	\begin{subequations}
		\begin{eqnarray}
			e_{00}^l &=& e_{0l} + e_{3l},\\
			e_{01}^l &=& e_{1l} - ie_{2l},\\
			e_{10}^l &=& e_{1l} + ie_{2l},\\
			e_{11}^l &=& e_{0l} - e_{3l},
		\end{eqnarray}
		\label{Eq:OldtoNewFormulation}
	\end{subequations}and
	\begin{subequations}
	\label{Eq:NewtoOldFormulation}
		\begin{eqnarray}
			e_{0l} &=& (e_{00}^l + e_{11}^l)/2,\\
			e_{1l} &=& (e_{01}^l + e_{10}^l)/2,\\
			e_{2l} &=& (e_{10}^l - e_{01}^l)/2i,\\
			e_{3l} &=& (e_{00}^l - e_{11}^l)/2.
		\end{eqnarray}
	\end{subequations}for $l=1,\ldots, N$.Thus, it is clear that one can
	describe a vector in the $\text{vec}$ formulation in terms of the
	coordinates of the vector in the $\zeta$ formulation.
	In order to show that these two formulations are equivalent, we need to show that the additive operation in one formulation
	can be described by the vectors in the other formulation.
	Let
	\begin{eqnarray}
		A &=& \sum_{i_1,j_1}\cdots \sum_{i_N,j_N} a^1_{i_1 j_1}\cdots a^N_{i_N j_N}\ketbra{i_1}{j_1}\cdots\ketbra{i_N}{j_N},\\
		B &=& \sum_{p_1,r_1}\cdots \sum_{p_N,r_N} b^1_{p_1 r_1}\cdots b^N_{p_N r_N}\ketbra{p_1}{r_1}\cdots\ketbra{p_N}{r_N}.
	\end{eqnarray}
	Then,
	\begin{eqnarray}
		\text{vec}(AB) &=& \sum_{p_1,r_1}\cdots \sum_{p_N,r_N} b^1_{p_1 r_1}\cdots b^N_{p_N r_N} (A\ket{p_1}\cdots\ket{p_N})\ket{r_1}\cdots\ket{r_N}\nonumber\\
					   &=& \sum_{i_1,\ldots, i_N} \sum_{r_1,\ldots,r_N} \Big{(}\sum_{p_1,\ldots,p_N}a^1_{i_1 p_1}b^1_{p_1 r_1}\cdots a^N_{i_N p_N}b^N_{p_N r_N}\Big{)}\ket{i_1}\cdots\ket{i_N})\ket{r_1}\cdots\ket{r_N}\nonumber\\
					   &=& \sum_{i_1,\ldots, i_N} \sum_{r_1,\ldots,r_N} \Big{[}(\sum_{p_1}a^1_{i_1 p_1}b^1_{p_1 r_1})\cdots (\sum_{p_N}a^N_{i_N p_N}b^N_{p_N r_N})\Big{]}\ket{i_1}\cdots\ket{i_N})\ket{r_1}\cdots\ket{r_N}.
	\end{eqnarray}We can describe each coordinate by
	\begin{equation}
		\text{vec}(AB)_{i_1,\ldots,i_N,r_1,\ldots,r_N} = (\sum_{p_1}a^1_{i_1 p_1}b^1_{p_1 r_1})\cdots (\sum_{p_N}a^N_{i_N p_N}b^N_{p_N r_N}).
	\end{equation}From Eq.~(\ref{Eq:OldtoNewFormulation}), denoting
	$\lambda_{i_2,\ldots,i_N,r_2,\ldots,r_N} = (\sum_{p_2}a^2_{i_2 p_2}b^2_{p_2 r_2})\cdots (\sum_{p_N}a^N_{i_N p_N}b^N_{p_N r_N})$,
	we obtain
	\begin{subequations}
	\label{Eq:VecZetaRelation}
		\begin{eqnarray}
			\text{vec}(AB)_{0,i_2,\ldots, i_N,0,r_2,\ldots,r_N} &=& [(a_{01} + a_{31})(b_{01} + b_{31}) + (a_{11} - ia_{21})(b_{11} + ib_{21})]\lambda_{i_2,\ldots,i_N,r_2,\ldots,r_N},\\
			\text{vec}(AB)_{0,i_2,\ldots, i_N,1,r_2,\ldots,r_N} &=& [(a_{01} + a_{31})(b_{11} - ib_{21}) + (a_{11} - ia_{21})(b_{01} - b_{31})]\lambda_{i_2,\ldots,i_N,r_2,\ldots,r_N},\\
			\text{vec}(AB)_{1,i_2,\ldots, i_N,0,r_2,\ldots,r_N} &=& [(a_{11} + ia_{21})(b_{01} + b_{31}) + (a_{01} - a_{31})(b_{11} + ib_{21})]\lambda_{i_2,\ldots,i_N,r_2,\ldots,r_N},\\
			\text{vec}(AB)_{1,i_2,\ldots, i_N,1,r_2,\ldots,r_N} &=& [(a_{11} + ia_{21})(b_{11} - ib_{21}) + (a_{01} - a_{31})(b_{01} - b_{31})]\lambda_{i_2,\ldots,i_N,r_2,\ldots,r_N}.
		\end{eqnarray}
	\end{subequations}Expanding $\lambda_{i_2,\ldots,i_N,r_2,\ldots,r_N}$ in terms of $a_{ij}$ and $b_{ij}$, we see that
	$\text{vec}(AB)$ can be computed from the vector representation given in Definition~\ref{Def:HomomorphismGN}. Similarly,
	Eq.~(\ref{Eq:NewtoOldFormulation}) can be applied in order to describe $\zeta(AB)$ in terms of $\text{vec}(A)$ and
	$\text{vec}(B)$.
\end{proof}

\begin{proposition}
	Let $\mathcal{S}$ be a stabilizer group with operators satisfying the structure of the standard stabilizer formalism. Assume that
	$\mathcal{C}$ is the additive group constructed using the standard stabilizer formalism and
	$\mathcal{C}_\text{vec} = \text{vec}(\mathcal{S})$, where the composition
	of operators in $\mathcal{S}$ corresponds to the respective operation of the additive group. Then
	$\mathcal{C} \equiv \mathcal{C}_\text{vec}$.
\end{proposition}

\begin{proof}
	Consider the single qubit $N = 1$ case. Let $A = X(a)Z(b)$, for $a,b\in\mathbb{Z}_2$. Then we can write
	\begin{equation}
		A = A_{00}\ketbra{0}{0} + A_{01}\ketbra{0}{1} + A_{10}\ketbra{1}{0} + A_{11}\ketbra{1}{1},
	\end{equation}where $A_{00} = 1 - a$, $A_{01} = (-1)^b a$, $A_{10} = a$, $A_{11} = (-1)^b(1 - a)$.
	These equalities are clearly invertible.
	Now, consider the case where $N>1$.	The coordinates of $\text{vec}(AB)$ are given by
	\begin{eqnarray}
		\text{vec}(AB)_{i_1,\ldots,i_N,r_1,\ldots,r_N} &=& (\sum_{p_1}a^1_{i_1 p_1}b^1_{p_1 r_1})\cdots (\sum_{p_N}a^N_{i_N p_N}b^N_{p_N r_N})\\
													   &=& AB_{i_1 r_1}^1\cdots AB_{i_N r_N}^N.
	\end{eqnarray}Then we can see that
	\begin{subequations}
		\begin{eqnarray}
			AB_{00}^j &=& (1 - a_1^j)(1 - (-1)^{b_1^j}) + (-1)^{b_1^j} a_2^j a_1^j,\\
			AB_{01}^j &=& (1 - a_1^j)(-1)^{b_1^j + b_2^j} + (1 - (-1)^{b_1^j})(-1)^{b_2^j} a_2^j a_1^j,\\
			AB_{10}^j &=& (1 - (-1)^{b_1^j})a_1^j + (-1)^{b_1^j} a_2^j(1 - a_1^j),\\
			AB_{11}^j &=& a_1^j (-1)^{b_1^j + b_2^j} + (1 - (-1)^{b_1^j})(-1)^{b_2^j}(1 - a_1^j)a_2^j.
		\end{eqnarray}
	\end{subequations}Since the above equalities are invertible, we have that both formulations are equivalent.
\end{proof}

\begin{example}
	Consider the operator $A = X\otimes \mathbb{I} + \mathbb{I}\otimes X$. The operator A can be
	written in the computational basis as
	\begin{eqnarray}
		A &=& (\ketbra{0}{1} + \ketbra{1}{0})\otimes(\ketbra{0}{0}+\ketbra{1}{1}) + (\ketbra{0}{0}+\ketbra{1}{1})\otimes(\ketbra{0}{1} + \ketbra{1}{0})\\
		  &=& \ketbra{0}{0}\otimes\ketbra{0}{1} + \ketbra{0}{1}\otimes\ketbra{0}{0} + \ketbra{0}{0}\otimes\ketbra{1}{0} + \ketbra{0}{1}\otimes\ketbra{1}{1}\nonumber\\
		  &+& \ketbra{1}{0}\otimes\ketbra{0}{0} + \ketbra{1}{1}\otimes\ketbra{0}{1} + \ketbra{1}{0}\otimes\ketbra{1}{1} + \ketbra{1}{1}\otimes\ketbra{1}{0}.
	\end{eqnarray}Thus, the vectorization of $A$ is given by
	\begin{eqnarray}
		\text{vec}(A) &=& \ket{0001} + \ket{0010} + \ket{0100} + \ket{0111}\nonumber\\
		  			  &+& \ket{1000} + \ket{1011} + \ket{1101} + \ket{1110}.
	\end{eqnarray}
\end{example}

\begin{example}
	In this example we show the equivalence between the vectorization
	representation and the $\zeta$-representation. Consider
	$A = X\otimes \mathbb{I}$ and $B = \mathbb{I}\otimes Z$. It is clear that
	$AB = X\otimes Z$. The $\zeta$ and $\text{vec}$ representations of $A$, $B$, and $AB$ are
	\begin{eqnarray}
	\zeta(A) &=& (0, 1, 1, 0, 0, 0, 0, 0),\\
	\text{vec}(A) &=& (0, 0, 1, 0, 0, 0, 0, 1, 1, 0, 0, 0, 0, 1, 0, 0),
	\end{eqnarray}
	\begin{eqnarray}
	\zeta(B) &=& (1, 0, 0, 0, 0, 0, 0, 1),\\
	\text{vec}(B) &=& (1, 0, 0, 0, 0, -1, 0, 0, 0, 0, 1, 0, 0, 0, 0, -1),
	\end{eqnarray}and
	\begin{eqnarray}
	\zeta(AB) &=& (0, 0, 1, 0, 0, 0, 0, 1),\\
	\text{vec}(AB) &=& (0, 0, 1, 0, 0, 0, 0, -1, 1, 0, 0, -1, 0, 0, 0, 0).
	\end{eqnarray}To show the correspondence between these two representations, we are going to describe
	$\text{vec}(AB)$ in terms of the elements in $\zeta(A)$ and $\zeta(B)$ using Eq.~(\ref{Eq:VecZetaRelation}).
	In particular, we have
	\begin{eqnarray}
		\text{vec}(AB)_{0,0,0,0} &=& [(a_{01} + a_{31})(b_{01} + b_{31}) + (a_{11} - ia_{21})(b_{11} + ib_{21})]\nonumber\\
							 &\times&[(a_{02} + a_{32})(b_{02} + b_{32}) + (a_{12} - ia_{22})(b_{12} + ib_{22})],\nonumber\\
		\text{vec}(AB)_{0,0,0,1} &=& [(a_{01} + a_{31})(b_{01} + b_{31}) + (a_{11} - ia_{21})(b_{11} + ib_{21})]\nonumber\\
							 &\times&[(a_{02} + a_{32})(b_{12} - ib_{22}) + (a_{12} - ia_{22})(b_{02} - b_{32})],\nonumber\\
		\text{vec}(AB)_{0,1,0,0} &=& [(a_{01} + a_{31})(b_{01} + b_{31}) + (a_{11} - ia_{21})(b_{11} + ib_{21})]\nonumber\\
							 &\times&[(a_{12} + ia_{22})(b_{02} + b_{32}) + (a_{02} - a_{32})(b_{12} + ib_{22})],\nonumber\\
		\text{vec}(AB)_{0,1,0,1} &=& [(a_{01} + a_{31})(b_{01} + b_{31}) + (a_{11} - ia_{21})(b_{11} + ib_{21})]\nonumber\\
							 &\times&[(a_{12} + ia_{22})(b_{12} - ib_{22}) + (a_{02} - a_{32})(b_{02} - b_{32})],\nonumber
	\end{eqnarray}
	\begin{eqnarray}
		\text{vec}(AB)_{0,0,1,0} &=& [(a_{01} + a_{31})(b_{11} - ib_{21}) + (a_{11} - ia_{21})(b_{01} - b_{31})]\nonumber\\
							 &\times&[(a_{02} + a_{32})(b_{02} + b_{32}) + (a_{12} - ia_{22})(b_{12} + ib_{22})],\nonumber\\
		\text{vec}(AB)_{0,0,1,1} &=& [(a_{01} + a_{31})(b_{11} - ib_{21}) + (a_{11} - ia_{21})(b_{01} - b_{31})]\nonumber\\
							 &\times&[(a_{02} + a_{32})(b_{12} - ib_{22}) + (a_{12} - ia_{22})(b_{02} - b_{32})],\nonumber\\
		\text{vec}(AB)_{0,1,1,0} &=& [(a_{01} + a_{31})(b_{11} - ib_{21}) + (a_{11} - ia_{21})(b_{01} - b_{31})]\nonumber\\
							 &\times&[(a_{12} + ia_{22})(b_{02} + b_{32}) + (a_{02} - a_{32})(b_{12} + ib_{22})],\nonumber\\
		\text{vec}(AB)_{0,1,1,1} &=& [(a_{01} + a_{31})(b_{11} - ib_{21}) + (a_{11} - ia_{21})(b_{01} - b_{31})]\nonumber\\
							 &\times&[(a_{12} + ia_{22})(b_{12} - ib_{22}) + (a_{02} - a_{32})(b_{02} - b_{32})],\nonumber
	\end{eqnarray}
	\begin{eqnarray}
		\text{vec}(AB)_{1,0,0,0} &=& [(a_{11} + ia_{21})(b_{01} + b_{31}) + (a_{01} - a_{31})(b_{11} + ib_{21})]\nonumber\\
							 &\times&[(a_{02} + a_{32})(b_{02} + b_{32}) + (a_{12} - ia_{22})(b_{12} + ib_{22})],\nonumber\\
		\text{vec}(AB)_{1,0,0,1} &=& [(a_{11} + ia_{21})(b_{01} + b_{31}) + (a_{01} - a_{31})(b_{11} + ib_{21})]\nonumber\\
							 &\times&[(a_{02} + a_{32})(b_{12} - ib_{22}) + (a_{12} - ia_{22})(b_{02} - b_{32})],\nonumber\\
		\text{vec}(AB)_{1,1,0,0} &=& [(a_{11} + ia_{21})(b_{01} + b_{31}) + (a_{01} - a_{31})(b_{11} + ib_{21})]\nonumber\\
							 &\times&[(a_{12} + ia_{22})(b_{02} + b_{32}) + (a_{02} - a_{32})(b_{12} + ib_{22})],\nonumber\\
		\text{vec}(AB)_{1,1,0,1} &=& [(a_{11} + ia_{21})(b_{01} + b_{31}) + (a_{01} - a_{31})(b_{11} + ib_{21})]\nonumber\\
							 &\times&[(a_{12} + ia_{22})(b_{12} - ib_{22}) + (a_{02} - a_{32})(b_{02} - b_{32})],\nonumber
	\end{eqnarray}
	\begin{eqnarray}
		\text{vec}(AB)_{1,0,1,0} &=& [(a_{11} + ia_{21})(b_{11} - ib_{21}) + (a_{01} - a_{31})(b_{01} - b_{31})]\nonumber\\
							 &\times&[(a_{02} + a_{32})(b_{02} + b_{32}) + (a_{12} - ia_{22})(b_{12} + ib_{22})],\nonumber\\
		\text{vec}(AB)_{1,0,1,1} &=& [(a_{11} + ia_{21})(b_{11} - ib_{21}) + (a_{01} - a_{31})(b_{01} - b_{31})]\nonumber\\
							 &\times&[(a_{02} + a_{32})(b_{12} - ib_{22}) + (a_{12} - ia_{22})(b_{02} - b_{32})],\nonumber\\
		\text{vec}(AB)_{1,1,1,0} &=& [(a_{11} + ia_{21})(b_{11} - ib_{21}) + (a_{01} - a_{31})(b_{01} - b_{31})]\nonumber\\
							 &\times&[(a_{12} + ia_{22})(b_{02} + b_{32}) + (a_{02} - a_{32})(b_{12} + ib_{22})],\nonumber\\
		\text{vec}(AB)_{1,1,1,1} &=& [(a_{11} + ia_{21})(b_{11} - ib_{21}) + (a_{01} - a_{31})(b_{01} - b_{31})]\nonumber\\
							 &\times&[(a_{12} + ia_{22})(b_{12} - ib_{22}) + (a_{02} - a_{32})(b_{02} - b_{32})].\nonumber
	\end{eqnarray}From the above relations and the representation $\zeta(A)$ and $\zeta(B)$, we see that the same result for
	$\text{vec}(AB)$ is obtained.
\end{example}

In the previous two examples we have seen vectorization applied to operators and that the $\zeta(\cdot)$ and $\text{vec}(\cdot)$ are indeed
equivalent. As can be noticed, the computations to implement the representations and to show equivalence are not complicate but tedious.
Therefore, such task can be delegated to a computer.

\subsection{Additive Codes}
\label{SubSec:AdditiveCodes}

As explained in the previous section, we need to have a symplectic form in order to construct the additive code
related to the stabilizer code and its centralizer. We can use Eq.~(\ref{Eq:CommutatorVec}) to construct the symplectic
form used through this section.

\begin{definition}
	Let $A,B\in\mathcal{L}(\mathcal{H}_2^{\otimes N})$ be linear operators. We define the map
	\begin{eqnarray}
		\langle\cdot, \cdot\rangle_{\text{vec}}\colon \mathbb{C}^{2N}\times\mathbb{C}^{2N}&\rightarrow& \mathbb{C},\\
		(\text{vec}(A),\text{vec}(B))&\mapsto& \langle\text{vec}(A),\text{vec}(B)\rangle_\text{vec} = \sum_{i=1}^{2N}[(A\otimes\mathbb{I} - \mathbb{I}\otimes A^T)\text{vec}(B)]_i.
	\end{eqnarray}
	\label{Def:SympFormVec}
\end{definition}

\begin{proposition}
	The map from Definition~\ref{Def:SympFormVec} is a symplectic form over $\mathbb{C}$.
\end{proposition}

\begin{proof}
	We show that the properties of Definition~\ref{Def:SympFormVec} are satisfied. Let
	$A,B,C\in\mathcal{L}(\mathcal{H}_2^{\otimes N})$ be operators. First of all, we see that
	\begin{eqnarray}
	\langle\text{vec}(A) + \text{vec}(B), \text{vec}(C)\rangle_\text{vec} &=& \sum_{i=1}^{2N}([(A+B)\otimes\mathbb{I} - \mathbb{I}\otimes (A+B)^T]\text{vec}(C))_i\nonumber\\
															   &=& \sum_{i=1}^{2N}[(A\otimes\mathbb{I} - \mathbb{I}\otimes A^T)\text{vec}(C)]_i + \sum_{i=1}^{2N}[(B\otimes\mathbb{I} - \mathbb{I}\otimes B^T)\text{vec}(C)]_i\nonumber\\
															   &=& \langle\text{vec}(A), \text{vec}(C)\rangle_\text{vec} + \langle\text{vec}(B), \text{vec}(C)\rangle_\text{vec}.
	\end{eqnarray}The second point follows from
	\begin{eqnarray}
		\langle\text{vec}(A), \text{vec}(B)\rangle_\text{vec} &=& \sum_{i=1}^{2N}[\text{vec}([A,B])]_i\\
													 &=& \sum_{i=1}^{2N}[\text{vec}(AB) - \text{vec}(BA)]_i\\
													 &=& -\sum_{i=1}^{2N}(\text{vec}(BA) - \text{vec}(AB))_i\\
													 &=& -\sum_{i=1}^{2N}[\text{vec}([B,A])]_i\\
													 &=& -\langle\text{vec}(B), \text{vec}(A)\rangle_\text{vec}.
	\end{eqnarray}We used the linearity of the vectorization in the second equality. The last point follows by expanding an
	operator $A$ in an eigenbasis and computing $(A\otimes\mathbb{I} - \mathbb{I}\otimes A^T)\text{vec}(A)$.
\end{proof}

Since the Definition~\ref{Def:SympFormVec} gives a symplectic form, we can define the dual code of an additive code.
Furthermore, we can extend the stabilizer formulation presented in the previous section to a larger set of errors.

\begin{definition}
	Let $C$ be an $+_\text{vec}$-additive code. The symplectic dual of $C$ is given by
	\begin{equation}
		C^{\perp_\text{vec}} := \{\bm{c}\in\mathbb{C}^{2N}\colon \langle \bm{c}, \bm{d}\rangle_\text{vec} = 0, \text{ for all }\bm{d}\in C\}.
	\end{equation}
\end{definition}

\begin{theorem}
	Let $\mathcal{V}_{\mathcal{S}_\text{DFS}} = \text{vec}(\mathcal{S}_\text{DFS})$ or
	$\mathcal{V}_{\mathcal{S}_\text{sDFS}} = \text{vec}(\mathcal{S}_\text{sDFS})$ be a basis of the
	$+_\text{vec}$-additive code $C$. Then,
		\begin{enumerate}
			\item A decoherence-free stabilizer code $\mathcal{Q}$ exists if there exists an $+_\text{vec}$-additive
			code $C$ over $\mathbb{C}$ generated by $\mathcal{V}_{\mathcal{S}_\text{DFS}}$ such that $C\leq C^{\perp_\text{vec}}$ and
			$\text{vec}(H_{ev})\in C^{\perp_\text{vec}}$;
			\item A strong decoherence-free stabilizer code $\mathcal{Q}$ exists if there exists an $+_\text{vec}$-additive code
			$C$ over $\mathbb{C}$ generated by $\mathcal{V}_{\mathcal{S}_\text{sDFS}}$ such that $C\leq C^{\perp_\text{vec}}$ and
			$\text{vec}(H_{S})\in C^{\perp_\text{vec}}$.
		\end{enumerate}
\end{theorem}

\begin{proof}
	It follows the same reasoning used in the proof of Theorem~\ref{Thm:StabilizerFormParticular}.
\end{proof}

\section{Application to Parameter Estimation}
\label{sec:channelEstimation}
In this section we consider the framework of quantum metrology and see how the stabilizer codes introduced previously can be used
as a tool within it. As it will be shown, one can reach the Heisenberg limit once the stabilizer codes attain an eigenvector condition.

Suppose we have an unitary evolution given by $U = \exp(-i H_S)$, where $H_S = \eta H$, $\eta$ is a parameter to be estimated, 
and $H$ is the generator of $U$.
One of the goals of quantum metrology is to reduce the variance obtained in estimating $\eta$ when compared to classical 
strategies. 
To attain this goal, we need to optimize the probing and measuring strategies. 
To decrease the estimating variance, we use $N$ identical and independent probes, measure them in the channel output, and average the results.
Such scheme has the estimation precision lower bounded by~\cite{Braunstein1996,Braunstein1994}
\begin{equation}
	\Delta\eta \Delta h\geq \frac{1}{2},
\end{equation}where $\Delta A$ is the variance of the random variable $A$, and $h = \sum_{j=1}^N H_j$, $H_j$ acting on the
$j$-th probe, stands for the generator of the unitary evolution $U^{\otimes N}$.
Furthermore, it is shown in Ref.~\cite{Giovannetti2006} that there exists a probing state and a measurement strategy such that 
\begin{equation}
	\Delta\eta \geq \frac{1}{N(\lambda_\text{Max} - \lambda_\text{Min})},
	\label{Eq:Giovannetti2006}
\end{equation}where $\lambda_\text{Max}$ and $\lambda_\text{Min}$ are the maximum and minimum eigenvalues of $h$. 
This is accomplished with the use of general probe states, which may be entangled states, and local or
joint measurements, after the unitary evolution $U^{\otimes N}$.
When the variance~(\ref{Eq:Giovannetti2006}) scales like $1/N$, we say that it attains the Heisenberg limit (HL) scaling.

A crucial assumption used in the above methodology to attain the HL is that the evolution is unitary. For Markovian noise,
one alternative approach is to use a quantum error-correcting code to achieve the HL under the assumption that the Hamiltonian is not in the spanned
space generated by the Lindblad
operators~\cite{Sekatski2017,DemkowiczDobrzanski2017,Zhou2018,Layden2019,Gorecki2020}.
Refs~\cite{Sekatski2017,DemkowiczDobrzanski2017} show that lower bounds can be constructed from a simple algebraic
condition involving solely the operators appearing in the quantum master equation.
a preliminary protocol considering the requirements that quantum error-correcting codes must satisfy in
order to achieve HL is also described in Ref.~\cite{DemkowiczDobrzanski2017}. This protocol has been extended
considering
necessary and sufficient conditions for achieving the HL when the probing system has a Markovian noise and
noiseless ancilla systems are available. This proposal has been further extended
for general adaptive multi-parameter estimation schemes in presence of Markovian noise~\cite{Gorecki2020}.
Lastly, Ref.~\cite{Layden2019} gives a semidefinite program for finding optimal ancilla-free sensing codes.

The proposed protocol of this paper is described as follows. The first part is the construction of the stabilizer code from the open quantum system evolution.
Let $\rho_{\text{Max-Min}}$ be the equally weighted superposition of the eigenvectors relative to the maximum and minimum eigenvalues of 
$\sum_{i=1}^N \mathbbm{1}^{\otimes i-1}_S\otimes H_S\otimes\mathbbm{1}^{\otimes N-i}_S$.
Next, we see if the stabilizer code contains the state $\rho_{\text{Max-Min}}$. If so, then we use it to probe the quantum system. 
As shown in Theorem~\ref{Theorem:DFStoStabilizer}, we are going to have an unitary evolution described by $H_S$. 
Therefore, using the optimal measurement described in Ref.~\cite{Giovannetti2006} over the channel outputs, one obtains the HL scaling.

\begin{figure}[h!]
	\begin{center}
		\includegraphics[width=.5\linewidth]{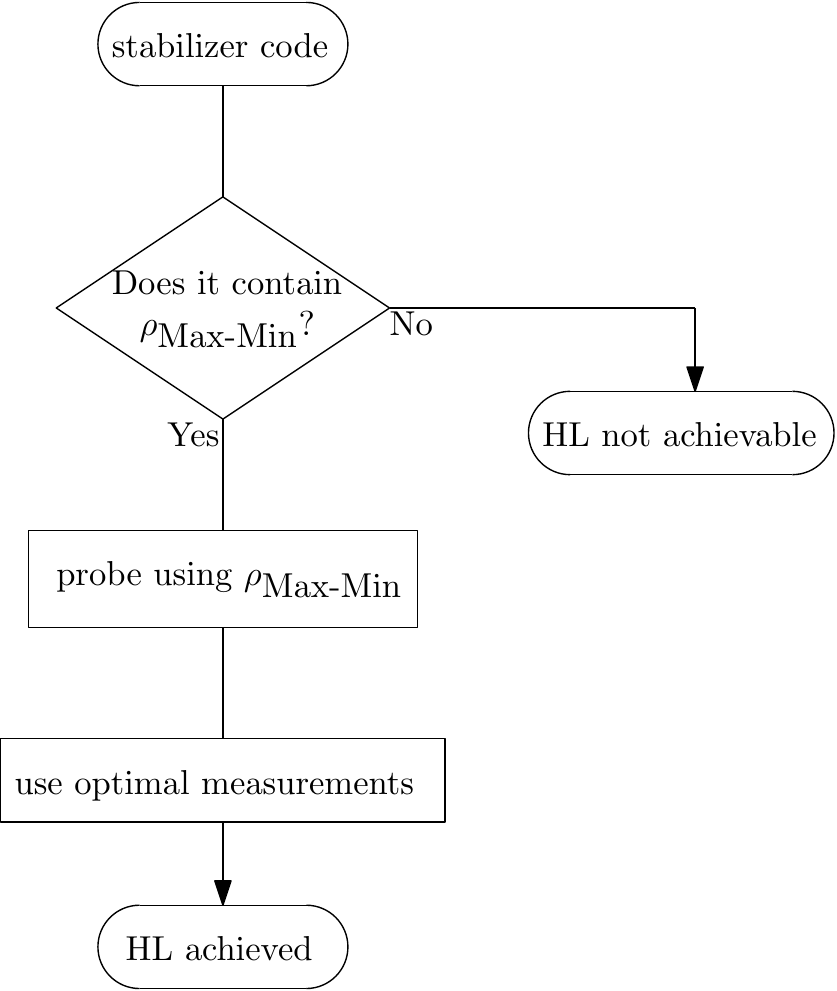}
	\end{center}
	\caption{Flowchart of the proposed protocol for achieving HL limit using the stabilizer codes of previous section.}%
	\label{fig:probingFlowchart}%
\end{figure}

The present idea differs from the literature on the use of quantum codes to attain the HL~\cite{Sekatski2017,DemkowiczDobrzanski2017,Gorecki2020,Zhou2018} in terms of computational
complexity.
Here, we do not need to implement a decoding process, which is the case of Refs.~\cite{Sekatski2017,DemkowiczDobrzanski2017,Zhou2018,Layden2019}.
However, this decoder-free approach is not novel in the literature, e.g. Ref.~\cite{Gorecki2020}
proposes a semidefinite program design to identify the optimal quantum error-correcting protocol,
without the need for a decoding algorithm, to achieve the best estimation precision in the case where the Heisenberg scaling is achievable.
The quantum state will not change by the environmental noise since it
belongs to the DFS. Therefore, there is no error to be detected or corrected. Removing the decoder from the picture, we have a reduced number of operations to be implemented
and a faster probing strategy.

\begin{theorem}
	Consider a quantum system with evolution given by a Markovian master equation with Lindblad operators $\{J_l\}$.
	Let $\mathcal{S}$ be a stabilizer set constructed from the Lindblad operators. Let $\ket{\psi_{\text{max}}}$ and
	$\ket{\psi_{\text{min}}}$ be eigenvectors of the system Hamiltonian $H_S$ with maximum and minimum eigenvalues,
	respectively. Then, Heisenberg limit scaling is achievable if
	\begin{equation*}
		\ket{\psi^{(N)}} = \frac{1}{\sqrt{2}}(\ket{\psi_{\text{max}}}^{\otimes N} + \ket{\psi_{\text{min}}}^{\otimes N})
	\end{equation*}belongs to the stabilizer code for any $N>N^*$, where $N^*\in\mathbb{N}$.
	\label{Theorem:HL}
\end{theorem}

\begin{proof}
	Since $\ket{\psi^{(N)}}$ belongs to the stabilizer code, then it also belongs
	to the DFS, hence its evolution is unitary and the technique of
	Ref.~\cite{Giovannetti2006} can be applied.
\end{proof}

We use Theorem~\ref{Theorem:HL} in the example below to show achievability of the HL
using the protocol of Fig.~\ref{fig:probingFlowchart}.
The proposed protocol relies on $\rho_{\text{Max-Min}}$ as a codeword
of the DFS stabilizer code. The existence of a DFS stabilizer code is equivalent to
the commutativity between the Lindblad operators and the system Hamiltonian.
This is satisfied whenever we have environments acting locally on each subsystem.
Therefore, we expect that the proposed protocol can be applied to most of the relevant
physical systems.

\begin{example}
	Consider a quantum system with a similar dynamics as in the previous examples,
      \begin{equation}
            \frac{\partial\rho}{\partial t} = -i[H_S,\rho] + \frac{\gamma}{2}(2J\rho J^\dagger - J^\dagger J\rho - \rho J^\dagger J),
            \label{Ex7}
      \end{equation}with
      \begin{equation}
            J = \frac{s+c}{2}(\mathbb{I}\otimes\mathbb{I} + \sigma_{z}\otimes\sigma_z),
      \end{equation}and 
      \begin{equation}
      		H_S = \frac{\gamma(s+c)^2}{4}(\sigma_{x}\otimes\sigma_x).
      \end{equation}where $s = \sinh(r)$, $c=\cosh(r)$, and $r$ is the (real) squeezing parameter. 
    The stabilizer set constructed from the dissipator part is given by $\mathcal{S} = \langle(\mathbb{I}\otimes\mathbb{I} + \sigma_{z}\otimes\sigma_z)^i\colon i=0,1\rangle$.
    Consider an eigenvector with maximum eigenvalue and an eigenvector with minimum eigenvalue of the operator $H_S$. Such a pair is
	\begin{equation}
		\ket{\psi_{\text{Max}}} = \frac{1}{\sqrt{2}}(\ket{00} + \ket{11}) \qquad \text{and} \qquad \ket{\psi_{\text{Min}}} = \frac{1}{\sqrt{2}}(\ket{00} - \ket{11}).
	\end{equation}Suppose we are going to probe the system $N$ times with the state
	\begin{equation}
		\rho_{\text{Max-Min}} = \ket{\psi^{(N)}}\bra{\psi^{(N)}} = \frac{1}{2}\Big{(}\ket{\psi_{\text{Max}}}^{\otimes N} + \ket{\psi_{\text{Min}}}^{\otimes N}\Big{)}\Big{(}\bra{\psi_{\text{Max}}}^{\otimes N} + \bra{\psi_{\text{Min}}}^{\otimes N}\Big{)}.
	\end{equation}It is possible to see that $\ket{\psi^{(N)}}$ is a codeword of the stabilizer code $\mathcal{Q}$, since $S\ket{\psi_{\text{Max}}} = \ket{\psi_{\text{Max}}}$ and $S\ket{\psi_{\text{Min}}} = \ket{\psi_{\text{Min}}}$, for all $S\in\mathcal{S}$.
	Now, the achievability of the HL scaling can by seen in two ways. Firstly from Theorem~\ref{Theorem:HL},
	where state membership in the stabilizer code is verified in the quantum or classical realms using the tools presented previously in this paper.
	Secondly from Eq.~(\ref{Ex7}), where we have that the dissipator part does not contribute to the evolution since 
      \begin{equation}
      2J\rho_{\text{Max-Min}} J^\dagger - J^\dagger J\rho_{\text{Max-Min}} - \rho_{\text{Max-Min}} J^\dagger J = 2\rho_{\text{Max-Min}} - \rho_{\text{Max-Min}} - \rho_{\text{Max-Min}} = 0.
      \end{equation}
      \label{Ex:HLscaling}
\end{example}

\section{Final Remarks}
\label{sec:Conclusion}
In this work we have constructed stabilizer codes for open quantum systems governed by Lindblad master equations. To achieve this goal, 
we had to go beyond the tools that exist for stabilizer codes in the literature. As an important step, we have extended the formulation
of stabilizer codes under the influence of errors forming a group to those forming a vector space.
Using stabilizer codes as tools,
we were able to determine conditions under which decoherence-free subspaces exist. As an application of the results shown, a novel
algebraic method for attaining the Heisenberg limit scaling is given by means of stabilizer codes. Explanations of tools and codes 
created in the paper are illustrated through a variety of examples. The algebraic approach developed to attain the Heisenberg limit scaling 
paves the way to attack this quantum metrology problem by reservoir engineering. 

This paper suggests future lines of investigation from coding theory perspective.
Firstly, by considering the construction of parameter bounds by connecting the physical constraints over Lindblad operators to the stabilizer code parameters.
A quantification of goodness for decoherence-free subspaces can be obtained from this topic. 
One could also show the non-existence of decoherence-free subspaces, which could lead to a more effective approach to investigate open quantum systems.
Secondly, identifying decoherence-free subspaces as stabilizer codes generates the possibility to classify some evolutions of open quantum systems.
One approach is connecting some evolutions to families of classical codes.
Lastly, because of the novel approach presented, we expect quantum evolutions with decoherence-free stabilizer codes
leading to classical codes that have not been discovered yet.


\section{Acknowledgments}
The authors acknowledge the funding from the European Union's Horizon 2020 research and innovation programme, 
under grant agreement QUARTET No 862644.

\addcontentsline{toc}{section}{References}
\nocite{*}
\printbibliography

\end{document}